\documentclass{aa}  

\usepackage{graphicx}
\usepackage{txfonts}
\usepackage{natbib}
\bibpunct{(}{)}{;}{a}{}{,}
\begin{document} 
\def\hi {H\,{\sc i}}
\def\hii {H\,{\sc ii}}
\def\water {H$_2$O}
\def\meth {CH$_{3}$OH}
\def\dg{$^{\circ}$} 
\def\kms{km\,s$^{-1}$}
\def\ms{m\,s$^{-1}$}
\def\jyb{Jy\,beam$^{-1}$}
\def\mjyb{mJy\,beam$^{-1}$}
\def\mjybcl{mJy\,beam$^{-1}$\,chan$^{-1}$}
\def\mjybc{$\rm{\frac{mJy}{beam\,chan}}$}
\def\mjybksl{mJy\,beam$^{-1}$\,km\,s$^{-1}$}
\def\mjybks{$\rm{\frac{mJy}{beam\,km\,s^{-1}}}$}
\def\solmass {\hbox{M$_{\odot}$}}
\def\solum {\hbox{L$_{\odot}$}} 
\def\d {$^{\circ}$}
\def\n {$n_{\rm{H_{2}}}$}
\def\kmsg{km\,s$^{-1}$\,G$^{-1}$}
\def\kmsM{km\,s$^{-1}$\,Mpc$^{-1}$}
\def\tbo {$T_{\rm{b}}\Delta\Omega$}
\def\tb {$T_{\rm{b}}$}
\def\om{$\Delta\Omega$}
\def\dvi {$\Delta V_{\rm{i}}$}
\def\dvz {$\Delta V_{\rm{Z}}$}
\def\code {FRTM code}
\def\NW {Nedoluha \& Watson}
\def\txs {TXS\,2226-184}
\title{VLBI observations of the \water ~gigamaser in \txs.}

\author{G.\ Surcis  \inst{1}
  \and 
  A. \ Tarchi \inst{1}
 \and
  P. Castangia \inst{1}
  }

\institute{INAF - Osservatorio Astronomico di Cagliari, Via della Scienza 5, I-09047, Selargius (CA), Italy\\
 \email{gabriele.surcis@inaf.it}
}  

\date{Received ; accepted}
\abstract
{Outside the Milky Way, the most luminous \water ~masers at 22~GHz, called 'megamasers' because of their extreme luminosity with
respect to the Galactic and extragalactic \water ~masers associated with star formation, are mainly detected in active galactic
nuclei. In the case of the \water ~maser detected in the nuclear region of the galaxy \txs 
~for the first time the term 'gigamaser' was used. However, the origin of this very luminous \water ~maser emission has never
been investigated into details.
}
{We study the nature of the 22~GHz \water ~gigamaser in \txs ~by measuring for the first time its absolute position at
milliarcsecond resolution, by comparing the morphology and characteristics of the maser emission at the very long baseline interferometry (VLBI) scales after about 
20 years, and by trying to detect its polarized emission.  
}
{We observed the 22~GHz \water ~maser emission towards \txs ~three times: the very first one with the NRAO Very Long Baseline
Array (VLBA, epoch 2017.45) and the next two times with the European VLBI Network (EVN, epochs 2017.83 and 2018.44). The first
two epochs (2017.45 and 2017.83) were observed in phase-reference mode, while the last epoch (2018.44) was observed in
full-polarization mode but not in phase-reference mode to increase the on-source integration time. We also retrieved and
analyzed the Very Long Baseline Array archival data at 22~GHz of \txs ~observed in 1998.40.}
{We detected six \water ~maser features in epoch 2017.45 (VLBA), one in epoch 2017.83 (EVN), and two in epoch 2018.44 (EVN).
All of them but one are red-shifted with respect to the systemic velocity of \txs, we detected only one blue-shifted maser
feature and it is the weakest one. All the \water ~maser features but the blue-shifted one are composed of two components with
very different linewidths. For the first time, we were able to measure the absolute position of the \water ~maser features with
errors below 1 milliarcsecond. No linear and circular polarization was detected.
}
{We were able to associate the \water ~maser features in \txs ~with the most luminous radio continuum clump reported in the
literature. The comparison between the epochs 1998.40 and 2017.45 reveals a difference in the morphology and velocity of the
maser features that can be justified accounting for maser variability. 
}
\keywords{Galaxies: masers - AGN }
\titlerunning{VLBI observations of the \water ~gigamaser \txs}
\authorrunning{Surcis et al.}

\maketitle
\section{Introduction}
\label{intro}
Water masers at 22 GHz are commonly detected at Galactic and extragalactic scales. In the Milky Way they are mainly detected
towards massive young stellar objects (YSOs), associated with molecular outflows. Outside our Galaxy most of them are
detected in active galactic nuclei (AGN), the so-called 'megamasers' ($\rm{L}>10$~\solum), and in starburst
galaxies, labeled as 'kilomasers' ($\rm{L}<10$~\solum)\footnote{Tarchi et al. (\citeyear{tar11}; their Sect. 4.2) warned
that this distinction should, however, be used with caution.}. \\
\indent \water ~masers associated with AGN have been related with three distinct phenomena: i) the nuclear accretion disk
(with the characteristic triple-peak pattern), where they can be used to trace the disk geometry and to measure the rotation
velocity and the enclosed nuclear mass (for recent cases see \citealt{gao17} and \citealt{zha18}); ii) the radio jets, 
where they
are the result of an interaction between the jet(s) and an encroaching molecular cloud (shocks) or due to an accidental
overlap along the line-of-sight between a cloud and the radio continuum of the jet, providing important information about
the evolution of jets and their hotspots (only a few sources of this kind have been studied in details: NGC\,1052: e.g.,
\citealt{cla98}; NGC\,1068: e.g., \citealt{gal01}; Mrk\,348: \citealt{pec03}); iii) the nuclear outflows, tracing the
velocity and geometry of nuclear winds at $<$ 1 pc from the nucleus (\citealt{gre03}, for the Circinus galaxy).\\
\indent Furthermore, by modeling the polarized emission of \water ~masers, it is possible to provide precise measurements of
the 3D geometry (linear polarization, P$_{l}$) and strength (circular polarization, P$_{V}$) of the magnetic field (e.g., 
\citealt{ned92}). For example, very long baseline interferometry (VLBI) studies of the \water ~maser polarization in the
Galactic YSO W3(H$_{2}$O) have revealed
extremely high $P_{l}$ percentage ($\sim40\%$; \citealt{god17}). However, so far only three extragalactic sources have been
investigated to detect polarized emission of \water ~maser: NGC\,4258 \citep{mod05}, NGC\,3079 \citep{vle07}, and the
Circinus galaxy \citep{mcc07}. No P$_{l}$ and P$_{V}$ were detected in any of these sources ($<1\%$).\\
\indent To date, more than 3000 galaxies have been searched for \water ~maser emission and detections have been obtained in
about 180 of them \citep{bra18}, the majority being radio-quiet AGN classified as Seyfert 2 (Sy 2) or low-ionization nuclear 
emission-line regions (LINERs), in the local Universe (z < 0.05). Notably, (almost) no \water ~masers have been found in
elliptical and/or radio-loud galaxies despite a number of surveys have been performed on different classes of AGN
\citep{tar12,hen18}. Nevertheless, there are few exceptions, for example the radio galaxy NGC\,1052 at z=0.005
\citep{bra94,cla98,saw08}, the FR\,II galaxy 3C403 at z=0.06 \citep{tar03,tar07}, the type 2 QSO\,SDSS J0804+3607 at
z=0.66 \citep{bar05,ben09}, and the type 1 QSO\,MG J0414+0534 at z=2.64 \citep{imp08,cas11}. \\
\begin {table*}[t]
\caption []{Observational details.} 
\begin{center}
\scriptsize
\begin{tabular}{ l c c c c c c c c c}
\hline
\hline
Array & observation     & Antennas                           & Source     &  \multicolumn{2}{c}{Pointing position}    & Integration    & Beam size          & Position & rms\tablefootmark{a}  \\ 
      & date            &                                    & name       & $\alpha_{2000}$  & $\delta_{2000}$        & time           &                    & Angle    &      \\ 
      &                 &                                    &            &($\rm{^{h}:~^{m}:~^{s}}$)&($\rm{^{\circ}:\,':\,''}$)& (hr)  &(mas~$\times$~mas)  & (\d)     & ($\frac{\rm{mJy}}{\rm{beam}}$) \\ 
\hline
VLBA  & 27 May 1998     & Br, Fd, Hn, Kp, La, Mk, Nl, Ov, Pt, Sc & \txs   & 22:29:12.494600 & -18:10:47.24200         &  7.16          & $1.81 \times 0.54$ & $-11.26$ & 5.9\tablefootmark{b}, \textbf{0.5}         \\
VLBA  & 12 June 2017    & Fd, Hn, Kp, La, Mk, Nl, Ov, Pt, Sc & \txs       & 22:29:12.494600 & -18:10:47.24200         &  4.13          & $1.42 \times 0.36$ & $-14.72$ & 15.3\tablefootmark{b}, \textbf{0.4}         \\
      &                 &                                    & J2232-1659\tablefootmark{c} & 22:32:22.564573  & -16:59:01.89226 & 4.21 & $1.34 \times 0.39$ & $-13.55$ & \textbf{1.6}                            \\
EVN   & 29 October 2017 & Ef, Mc, On, Ys, Hh                 & \txs       & 22:29:12.494600 & -18:10:47.24200         &  2.95          & $4.16 \times 0.97$ & $-26.99$ & 21.7\tablefootmark{b,d}, \textbf{0.2}      \\
      &                 &                                    & J2232-1659\tablefootmark{c} & 22:32:22.564573  & -16:59:01.89226 & 3.04 & $5.86 \times 0.92$ & $-30.28$ & \textbf{1.5}\tablefootmark{e}          \\
EVN   & 09 June 2018    & Ef, Mc, On, Ys, HH, Sr, Jb         & \txs       & 22:29:12.494600 & -18:10:47.24200         &  3.79          & $2.11 \times 1.41$ & $+7.08$ & 5.1\tablefootmark{b}, \textbf{0.2}        \\
\hline
\end{tabular}
\end{center}
\tablefoot{
\tablefoottext{a}{The rms obtained for the radio continuum by averaging the channels that do not show any maser emission is
indicated in boldface.}
\tablefoottext{b}{Measured in channels with no line emission.}
\tablefoottext{c}{Phase-reference source at 1.4\d ~from \txs. The errors of $\alpha_{2000}$ and $\delta_{2000}$ are 0.291~mas and 0.409~mas, respectively (see Goddard Space Flight Center VLBI group, February 23, 2015$^3$). }
\tablefoottext{d}{The measured rms per channel was 5 mJy/beam that needs to be multiplied by the estimated flux correcting
factor of 4.35 (see Sect.~\ref{evn_obs}).}
\tablefoottext{e}{The measured rms was 0.34 mJy/beam that needs to be multiplied by the estimated flux correcting factor of 4.35 (see Sect.~\ref{evn_obs}).}
}
\label{Obs}
\end{table*}
\indent Although the limited number of detections do not allow us to deduce any conclusive assertion, we can point out that
the detection of masers in E/radio-loud objects seems not to be particularly AGN-type dependent. Indeed, so far, \water 
~masers have been detected both in type\,1 and type\,2 objects. In general, \water ~masers in E/radio-loud objects are more
likely associated with radio-jets or outflows, although the disk-maser scenario cannot be a-priori ruled out. Obviously, VLBI
observations are necessary to confidently determine the origin of the emission. However, these measurements have been
successfully performed only for the nearest source, NGC\,1052 (at 20~Mpc distance), due to the intrinsic weakness or
flaring-down of the maser lines in the other cases. In NGC\,1052, high resolution observations indicate that the maser clouds
are most likely located foreground to the jet in a circumnuclear torus (or disk), thus amplifying the continuum seed emission
from the jet knots \citep{saw08}. Recently, the detection and interferometric follow-up with the Very Long Baseline Array
(VLBA) of a megamaser in the lenticular S0 galaxy IRAS15480-0344 (130~Mpc distance) has been reported by \citet{cas16}. Also
in this case, an association of (part of) the \water ~maser emission with a nuclear jet or outflow is proposed \citep{cas19}.
At a similar distance to that of IRAS15480-0344, an obvious and interesting case is represented by \txs ~(see next section
for details).
\section{The galaxy TXS\,2226-184}
\label{TXS}
\txs, located at distance of 107.1~Mpc (derived assuming $H_{\rm{0}}=70$~\kmsM, \citealt{kuo18}) and with a
systemic velocity of $V_{\rm{lsr, radio}}=7270$~\kms ~\citep{tay02}, has been optically classified as an elliptical/S0
galaxy \citep{koe95}, even though \citet{fal00} proposed an alternative classification as a possible later-type galaxy.
Furthermore, \txs ~is spectroscopically identified as a LINER \citep{ben04}, which is a signature of an AGN or a nuclear
starburst. Indeed, despite 
the radio core is relatively weak (or heavily obscured), the symmetric and tight structure of the nuclear radio 
emission is seemingly produced by a jet (position angle $\rm{PA}=-36$\d) that extends over $\sim100$~pc oriented at
large angle to the line of sight \citep{tay04}.\\
\indent \txs ~hosts one of the most luminous known extragalactic \water ~maser sources, the so called 'gigamaser',
with an isotropic luminosity of 6100~\solum ~\citep{koe95}. The maser line is broad ($\sim90$~\kms) and the bulk of
the emission is red-shifted ($V_{\rm{lsr, radio}}=7300-7500$~\kms) with respect to the velocity of the galaxy
\citep{koe95}. In addition, the observations of the \water ~gigamaser over a period of years revealed an uncommon 
stability of the emission (e.g., \citealt{bra03}). Because of the extreme brightness of the \water ~maser emission and of its broad
width, \citet{tay04} suggest, among other options, that the gigamaser might be associated with a
jet that drives shocks into a molecular cloud,
this is supported by the distribution of the \hi ~absorption that they detected towards the center of the galaxy.\\
\indent \citet{bal05} observed the gigamaser with the VLBA and identified seven
clumps of \water ~maser emission that are mainly distributed linearly northeast to southwest with $\rm{PA=+25}$\d. 
They also detected two blue-shifted maser clumps: one along the linear distribution (the most
southwest clump) and the other one, which is the bluest one ($V_{\rm{lsr, radio}}=7100$~\kms), at about 7~mas
($\sim$3.5~pc) southeast from the linear distribution. They also reported that in two of the maser clumps they detected 
a systemic position-velocity gradient. The observations were not conducted in phase-reference mode, 
therefore \citet{bal05} were not able to provide absolute positions of the maser clumps. Furthermore, no information on
the absolute positions is available in the literature so far. \citet{bal05} associate the linearly distributed maser
clumps with a parsec-scale, edge-on, rotating disk, whereas the isolated blue-shifted clump might be associated with a
jet. In addition, no 22~GHz continuum emission was detected by \citet{bal05}.\\
\indent To further investigate the nature of the \water ~gigamaser in \txs ~by providing the absolute position of
the maser, and to compare the spatial distribution of the maser features (clumps) after about 20 years, we decided
to observe \txs ~again with the VLBA (one epoch) and the European VLBI Network (EVN; two epochs) at 22~GHz. While the
VLBA epoch and the first EVN epoch were observed in phase-reference mode in order to measure the absolute positions of the
maser features, the second EVN epoch was performed in full polarization mode in order to attempt the detection of
polarized maser emission.  We report the VLBA and EVN observations and the corresponding analysis in Sect.~\ref{obssect},
including the VLBA archival data of \citet{bal05}. The results are instead described in Sect.~\ref{res}
and discussed in Sect.~\ref{discussion}.
\section{Observations and analysis}
\label{obssect}
In the following we will describe the observations, the calibration, and the analysis of the four different VLBI epochs: 
the first one is the VLBA archival data (epoch 1998.40) and then the new VLBA epoch and the two EVN
epochs, with the latest one observed in full polarization mode. In addition, we will also describe the archival data of 
the single-dish observations made with the Green Bank Telescope (GBT) to which we will refer in our discussion.
\subsection{\object{Archival data: VLBA data epoch 1998.40}}
\label{vlba_arch}
In order to recover more details than those reported in \citet{bal05} and to compare with the results of our recent
observations (see Sects.~\ref{vlba_obs}, \ref{evn_obs}, and \ref{evn_obspol}), we retrieved from the National Radio Astronomy Observatory (NRAO)\footnote{The NRAO is a
facility of the National Science Foundation operated under cooperative agreement by Associated Universities, Inc.}
data archive the VLBA dataset of the project BG0082 presented in \citet{bal05}.\\
\indent The 22 GHz \water ~maser emission (rest frequency: 22.23508~GHz) was observed towards \txs ~(named
IRAS\,22265 in BG0082) with ten antennas of the VLBA for a total observing time of 8.65 h on May 27, 1998. The
observing setup consisted of eight single-polarization subbands of 8~MHz, six of which were contiguous: subbands
from 1 to 6 were set in left circular polarization (LCP) and subbands 7 and 8 in right circular polarization (RCP);
in addition, subbands 2 and 4 were centered at the same frequencies of subbands 3 and 5, respectively. The total
covered velocity ($V_{\rm{lsr,radio}}$) range is of 650~\kms, i.e., from 6942~\kms ~to 7592 \kms. Each
subband was correlated with 256 channels, implying a channel width of $\sim$30~kHz (0.45~\kms). For more
observational details see Table~\ref{Obs}.\\
\indent The data were calibrated using the Astronomical Image Processing System (AIPS). Since the data are from the
90s, some preliminary steps were necessary before proceeding with the standard calibration applied nowadays. We
had to pre-process the external VLBA calibration files bg082cal.vlba and vlba\_gains.key by using the AIPS task
VLOG. This task creates external files needed for generating the AIPS tables TY, FG, and GC that contain
information on system temperatures (tsys), flags, and gains, respectively. We first performed the a-priori calibration
that consists of parallactic angle and amplitude corrections, then the bandpass, the delay, rate, and phase calibrations
were performed on the fringe-finder calibrator 3C454.3 (peak flux density~=~3.9~\jyb; $\rm{rms}=18.7$~\mjyb) that
was observed for 5 minutes every $\sim$25 minutes. Before shifting the frequency with the AIPS task CVEL to take
the rotation of the correlation center into account, we applied all the calibration solutions to \txs.
Because the \water ~maser was bright enough we decided to self-calibrate the peak flux channel and transfer the 
phase solutions to all channels and subbands. Before producing the final image cube we glued together, by using the
AIPS task UJOIN, the subbands 1, 2+3, 4+5, 6, 7 and 8, where 2+3 indicates the average of the identical 
subbands 2 and 3, and 4+5 of the identical subbands 4 and 5. From the image cube produced using the AIPS task 
IMAGR, we identified the 22~GHz \water ~maser features using the process described in \citet{sur11} considering
a maser detection threshold of $3\sigma$. No phase-reference source was observed, hence no absolute position estimation
was possible.
\subsection{\object{Archival data: GBT data epoch 2010.94}}
\label{vlba_arch}
To compare the single-dish profile of the 22~GHz \water ~maser with the profiles of the \water ~maser features  detected 
with the interferometers used in this work, we retrieved a dataset from the GBT archive. This was observed on December 9, 
2010, under project name AGBT10C\_013. Observations were performed in total power nod mode, using two of the seven beams of the
K-band focal plane array (KFPA) receiver with dual circular polarization. The spectrometer was configured with two 200~MHz 
IFs, each with 8192 channels, yielding a channel spacing of 24~kHz, corresponding to $\sim$0.3~\kms ~at the frequency of 
22~GHz. The first spectral window was centered at the frequency corresponding to the recessional velocity of the galaxy 
and the second was offset by 180~MHz to the red, for a total frequency coverage of 380\,MHz ($\sim$5100~\kms ~at 22~GHz). 
The data were reduced and analyzed using GBTIDL\footnote{http://gbtidl.nrao.edu/}. We calibrated the spectra utilizing 
standard routines and applying the default zenith opacity and gain curve. The uncertainty in this flux calibration 
procedure is estimated to be no more than 20\%. In order to obtain the final spectrum, we averaged the two polarizations 
and then subtracted a polynomial baseline of one degree from the spectrum.
\subsection{\object{New observations: VLBA data epoch 2017.45}}
\label{vlba_obs}
We observed the $6_{16}-5_{23}$ \water ~maser transition towards \txs ~with nine antennas of the VLBA on June 12,
2017. The observations were conducted in phase-reference (with cycles phase-calibrator -- target of 45~sec -- 
45~sec) and spectral mode for a total observing time, including overheads, of 10~h. We did
not include geodetic blocks in our observations. Indeed, these are fundamental to correct the
slowly varying term of the tropospheric delay and to improve the astrometric accuracy up to $\rm{\mu as}$ 
(for more details see \citealt{rei14}). However, our scientific purpose does not request an astrometric accuracy
better than 0.1~mas.
We used one dual-polarization baseband of 64~MHz centered at
the systemic velocity of the galaxy ($V_{\rm{lsr,radio}}=7270$~\kms) to cover a velocity range of about 870~\kms. A
single DiFX correlation pass was performed with 4096 channels that provided a velocity resolution of 0.24~\kms
~(channel width $\sim$16~kHz). For more observational details see Table~\ref{Obs}.\\
\indent The data were calibrated using AIPS. We first performed the a-priori calibration that consists of
ionospheric, earth rotation, parallactic angle, and amplitude corrections. The bandpass calibration was performed
on the phase-reference source J2232-1659, while the delay and phase calibrations were performed on the
fringe-finder 3C345. Then we self-calibrated the phase-reference source J2232-1659 (peak flux density~$=
145.9$~\mjyb; $\rm{rms}=1.6$~\mjyb) and the final solutions were transferred to the \txs ~data before shifting the
frequency with the AIPS task CVEL. Note that the uncertainties for the right ascension ($\alpha_{2000}$) and 
declination ($\delta_{2000}$) of the absolute position of J2232-1659 are 0.291~mas and 0.409~mas, 
respectively (see Goddard Space Flight Center VLBI group, February 23, 2015\footnote{http://www.aoc.nrao.edu/software/sched/catalogs/sources.gsfc}).
We self-calibrated an average of 20 channels around the peak of the
brightest maser feature (7404.4~\kms) to improve the phase solutions, this does not affect the application of the
solutions transferred from J2232-1659. Indeed a comparison of the brightest maser feature before and after applying the
self-calibrated solutions show no position difference. Before producing the final
image cube with the AIPS task IMAGR (field $0.2~\rm{arcsec}\times0.2~\rm{arcsec}$), we subtracted the continuum in the 
uv-plane using the AIPS task UVLSF by least-squares fitting the baseline in the free-line channels. We identified 
the 22~GHz \water ~maser features using the process described in \citet{sur11} (detection threshold
$>3\sigma$). We determined the absolute positions of the identified maser features by using the AIPS task JMFIT.
The Gaussian fit error of JMFIT is of the order of 0.1--0.4~mas depending on the
maser feature; in addition the uncertainty due to the thermal noise calculated before applying the self-calibration
\citep{rei14} was of the order of 0.01--0.05~mas.
Since we did not correct the slow term of the tropospheric delay, we decided to consider conservative position errors
equal to half the beam projected along $\alpha_{2000}$ and $\delta_{2000}$, which are $\pm0.2$~mas and $\pm0.7$~mas,
respectively.\\
\indent We made a continuum map to attempt the detection of any 22~GHz continuum emission of the nuclear region of \txs 
~by averaging the channels that do not show any maser emission.
\begin {table*}[t]
\caption []{Parameters of the 22 GHz \water ~maser features detected towards \txs ~in epoch 1998.40. The columns are: the
maser feature name, right ascension offset, declination offset, identified component from the Gaussian fit, peak flux density,
peak velocity (radio convention), linewidth of the maser feature, spectral noise per channel, signal-to-noise ratio, 
spectral noise per \kms, integrated flux, isotropic luminosity.} 
\begin{center}
\scriptsize
\begin{tabular}{ l c c c c c c c c c c c}
\hline
\hline
\,\,\,\,\,(1)&(2)                     & (3)                                     & (4)       & (5)        & (6)           &(7)                                   &(8)                       &(9)  & (10) & (11) & (12)\\
Maser     & $\alpha$\tablefootmark{a} & $\delta$\tablefootmark{a}& Component & Peak flux\tablefootmark{b}  & $V_{\rm{lsr,radio}}$\tablefootmark{b}& $\Delta v\rm{_{L}}$\tablefootmark{b} & rms\tablefootmark{b,c}     & SNR & rms & $ \int S dV$\tablefootmark{b} & $L_{\rm{H_2O}}$\tablefootmark{d}\\
          &  offset                   &     offset               &           &  Density(I) &               &                     &         &  & & &\\
          & (mas)                     & (mas)                    &           & (\mjyb)    &      (\kms)   &      (\kms)         & (\mjybc) &  & (\mjybks) &(Jy \kms) &($L_{\odot}$)\\ 
\\
\hline
TXS.B01   &     0                     &  0                        &  1   & $36.37$ &  $7395.5\pm0.7$  &  $61.8\pm2.6$  & 5.3 & 14\tablefootmark{e}& 3.43 & $2.4\pm0.1$ & $869$\tablefootmark{e}\\
          &                           &                           &  2   & $38.83$ &  $7391.6\pm0.3$  &  $20.8\pm1.7$  &     &                      & & $0.9\pm0.1$ &\\
TXS.B02   &    +0.601                 &   +0.800                  &  3   & $13.35$ &  $7391.3\pm1.4$  &  $39.0\pm5.3$  & 5.7 & 28\tablefootmark{e}& 3.69&$0.5\pm0.1$ & $632$\tablefootmark{e}\\ 
          &                           &                           &  4   & $14.39$ &  $7355.3\pm4.0$  &  $121.3\pm5.1$ &     &                      & & $1.9\pm0.1$ & \\ 
TXS.B03   &    +1.602                 &    +3.200                 &  5   & $37.66$ &  $7354.2\pm0.6$  &  $25.9\pm1.7$  & 5.1 & 7  & 3.31 & $1.39\pm0.04$ & $366$\\ 
\hline
\end{tabular} \end{center}
\tablefoot{
\tablefoottext{a}{No absolute positions were measured.}
\tablefoottext{b}{Gaussian fit results.}
\tablefoottext{c}{The channel width is 0.42~\kms.}
\tablefoottext{d}{Isotropic luminosities are derived from $L_{\rm{H_{2}O}}/[\solum]=0.023 \times \int S \, \rm{d}\textit{V}/[\rm{Jy \ km \ s^{-1}}] \times \textit{D}^{2}/[\rm{Mpc^{2}}]$, where $D=107.1$~Mpc (derived assuming $H_{\rm{0}}=70$~\kmsM, \citealt{kuo18}).}
\tablefoottext{e}{Estimated considering the sum of the two components.}
}
\label{1998_tab}
\end{table*}
\begin{figure*}[th!]
\centering
\includegraphics[width = 6 cm]{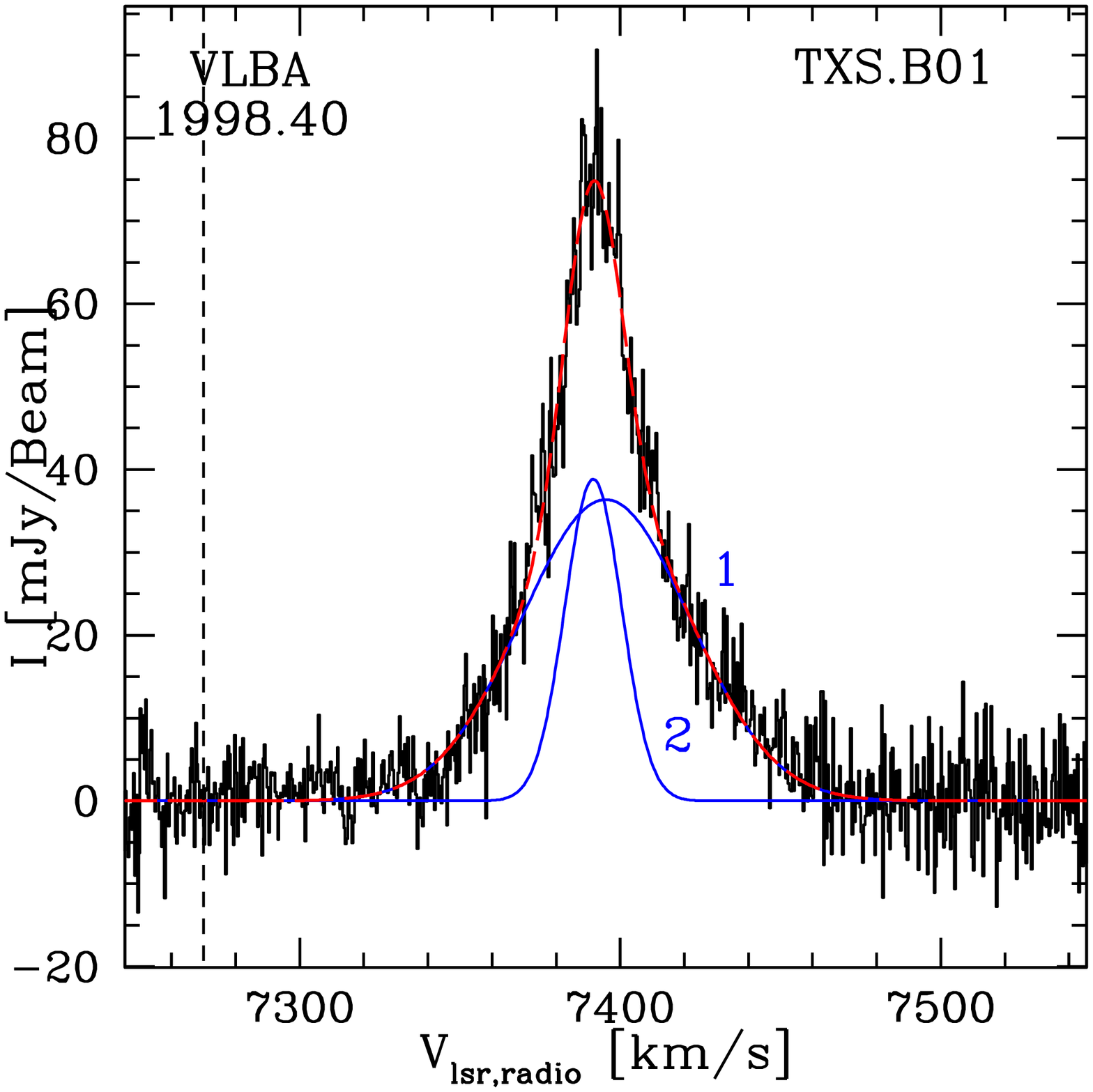}
\includegraphics[width = 6 cm]{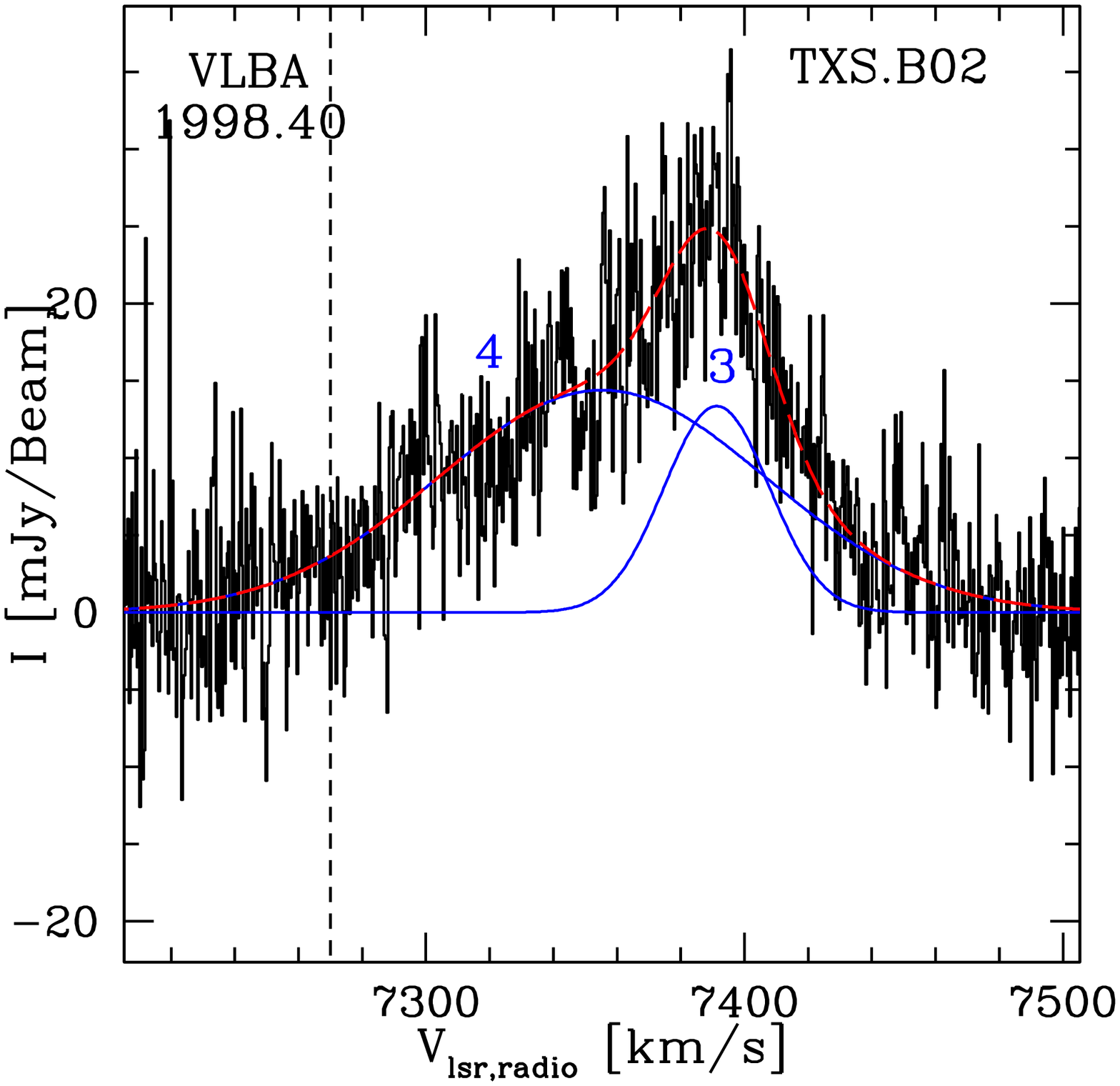}
\includegraphics[width = 6 cm]{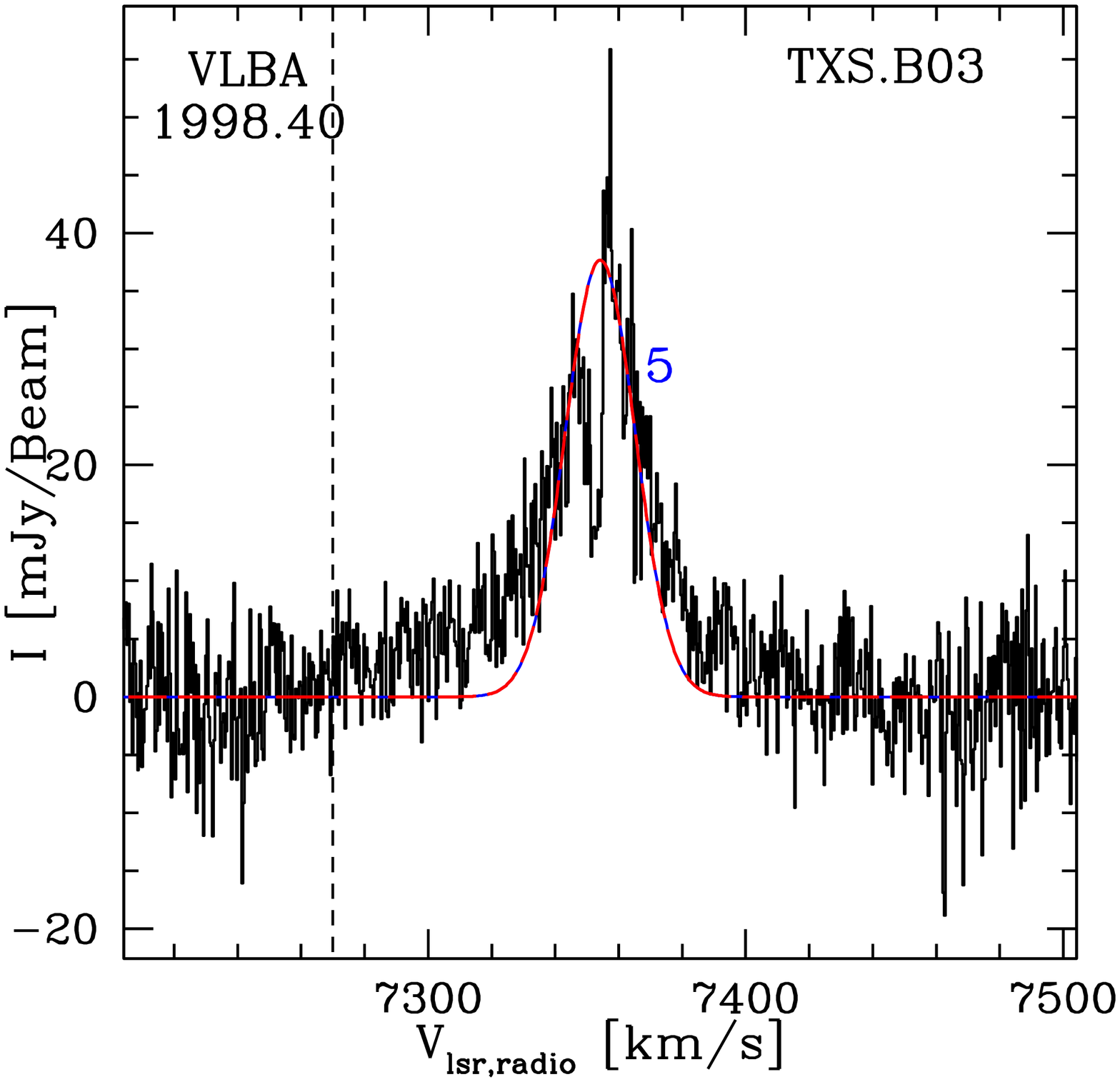}
\caption{Total intensity spectra ($I$) of the \water ~maser features detected towards \txs ~with the VLBA in epoch
1998.40. The thick blue lines are the Gaussian fit of the different components that are labelled in blue (see
Table~\ref{1998_tab}), the dashed red line is the combination of the Gaussian components for each maser
feature. The dashed black line indicates the systemic velocity of the galaxy in the radio convention
(i.e., 7270~\kms).}.
\label{vlba_1998}
\end{figure*}
\begin{figure*}[ht!]
\centering
\includegraphics[width = 9 cm]{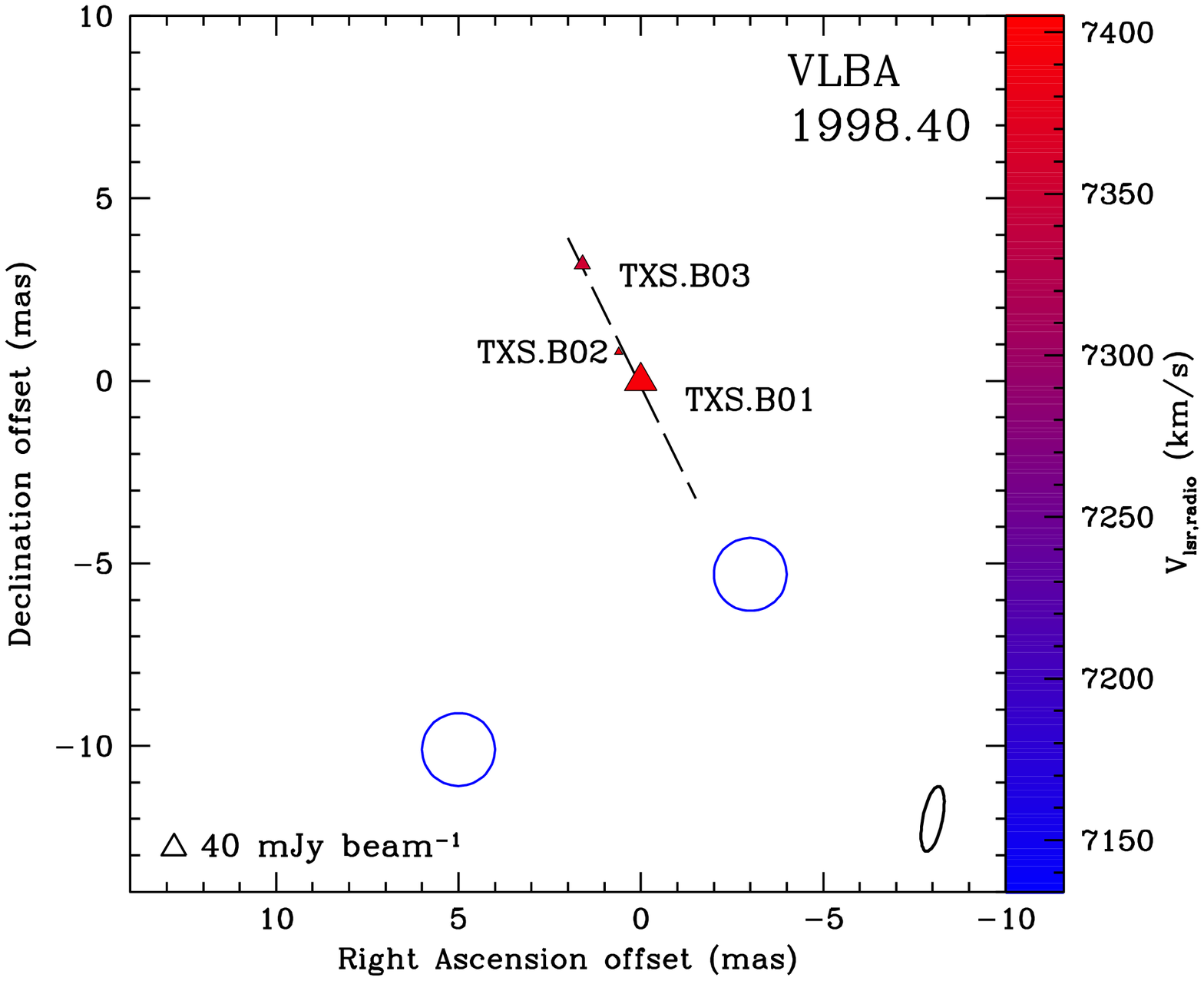}
\includegraphics[width = 9 cm]{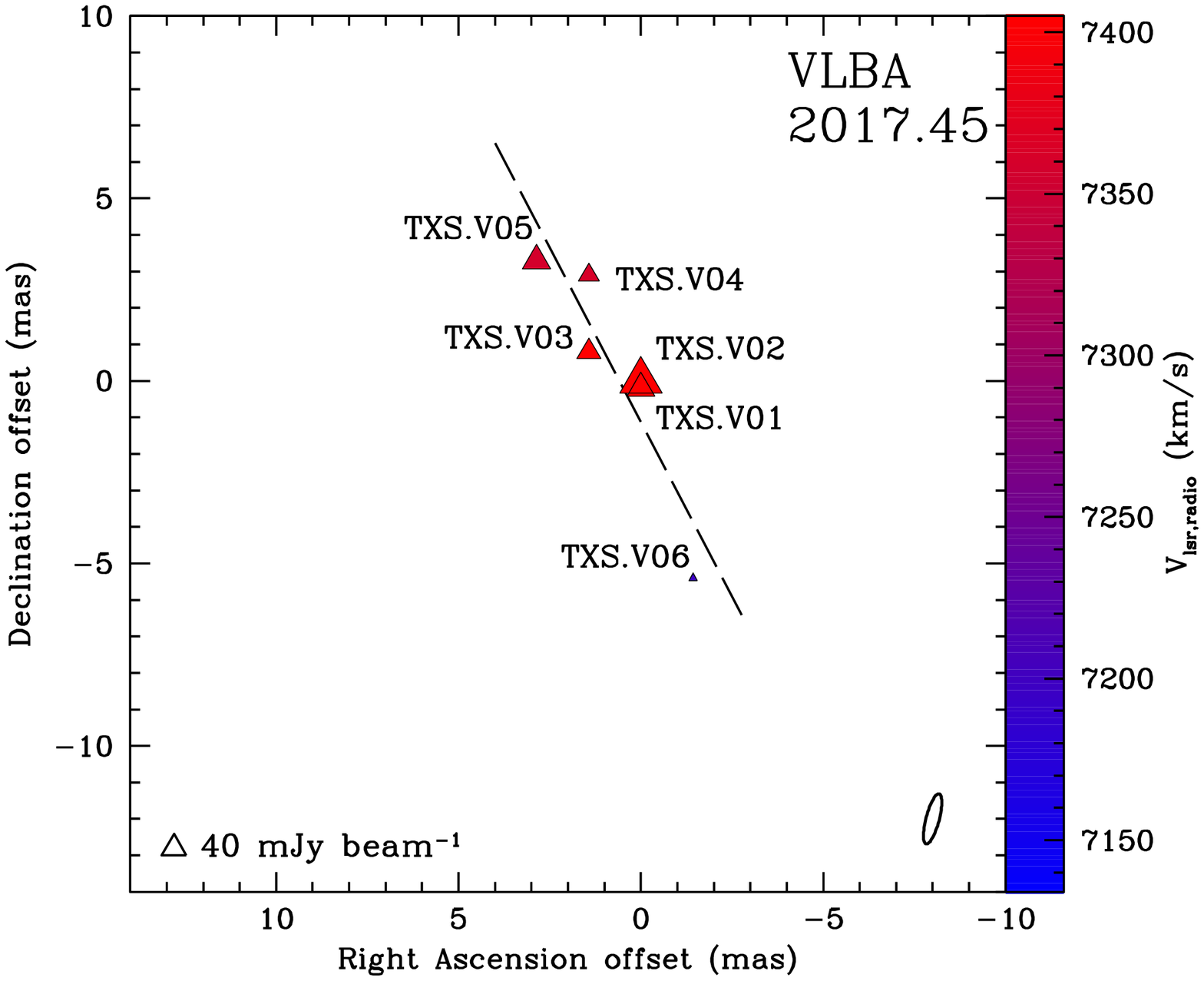}
\caption{View of the \water ~maser features detected towards \txs ~with the VLBA in epoch 1998.40 (\textit{left 
panel}) and in epoch 2017.45 (\textit{right panel}). Triangles identify \water ~maser features whose side length 
is scaled logarithmically according to their peak flux density (Table~\ref{1998_tab} and \ref{2017_tab}). 
40~\mjyb ~symbol is plotted for comparison in both panels. Maser local standard of rest radial velocities are
indicated by color (the velocity of the galaxy is $V_{\rm{lsr,radio}}^{\rm{TXS}}=7270$~\kms). The beam size (see Table~\ref{Obs})
is shown on the bottom right corner. At the distance of the galaxy (i.e, $107.1$~Mpc, derived assuming 
$H_{\rm{0}}=70$~\kmsM, \citealt{kuo18}) 1~mas corresponds to $\sim0.5$~pc.
\textit{Left panel}: in absence of absolute position measurements the relative positions of all maser features are
evaluated considering the brightest maser feature (TXS.B01) as reference. The two blue circles indicate the area
where \citet{bal05} detected the blue-shifted maser features and we did not (detection threshold of $3\sigma$). 
The dashed line is the best least-squares linear fit of the maser distribution
($\rm{PA_{H_{2}O}}=+26^{\circ}\pm16^{\circ}$). \textit{Right panel}: The reference position is the absolute position of
TXS.V02 (see Table~\ref{2017_tab}). The dashed line is the best least-squares linear fit of the maser distribution 
($\rm{PA_{H_{2}O}}=+28^{\circ}\pm12^{\circ}$).}.
\label{posplot_vlba}
\end{figure*}
\subsection{\object{New observations: EVN data epoch 2017.83}}
\label{evn_obs}
The \water ~maser transition at 22.23508~GHz was also observed with five of the EVN\footnote{The European VLBI Network
is a joint facility of European, Chinese, South African and other radio
astronomy institutes funded by their national research councils.} antennas on October 29, 2017. Similarly to
the VLBA observations (see Sect.~\ref{vlba_obs}), we performed phase-referenced spectral observations with cycles 
phase-calibrator -- target of 45~sec -- 45~sec. Similarly to the VLBA observations of epoch 2017.45, we did 
not include geodetic blocks. Here, we used two subbands of 32~MHz each (435~\kms) with an overlap
between them of 10~MHz to avoid problems with the edges of the subbands. Therefore, we covered a total velocity
range of $\sim650$~\kms ~that is enough to fully detect the broad maser emission ($\sim200$~\kms). The total
observation time including the overheads was 7~h. The data were correlated with the EVN software correlator (SFXC,
\citealt{kei15}) at the Joint Institute for VLBI ERIC (JIVE) using 2048 channels per subband
(channel width $\sim$16~kHz~$\sim0.2$~\kms ~at 22~GHz). Further observational details are reported in Table~\ref{Obs}.
During our observations an issue with the Digital Base Band Converter (DBBC) firmware affected the participating antennas, in case of
2~Gbps recording and/or 32~MHz subbands, leading to an erroneous, very low amplitudes (e.g., \citealt{bur19}).\\
\indent The data were calibrated with AIPS. After applying the a-priori calibration obtained from the JIVE
pipeline, we performed the bandpass calibration and the fringe-fitting on the calibrators J2232-1659 and 3C454.3,
respectively. We self-calibrated the phase-reference source J2232-1659 that showed a peak flux density of 
33.5~\mjyb ~and a continuum rms of 0.3~\mjyb, which are much lower than those measured with the VLBA. We note that
the theoretical noise between the two observations are identical within 10\%. However, this low amplitudes are
expected, as mentioned above, because of the technical issue encountered during the EVN session 3/2017. Therefore,
to correct the EVN amplitudes we assumed that the flux of J2232-1659 was constant between June and October
2017. This leads us to determine that all the amplitudes in the EVN dataset need to be multiplied by a
factor of 4.35, similar to what was estimated by another EVN projects observed during the same session
(R.A.~Burns, private communication). Therefore, we measured for 3C454.3 a corrected peak flux of 7.9~\jyb. \\
\indent With the AIPS task CVEL we shifted the frequency after applying the phase solutions from the 
phase-reference calibrator to \txs. Before producing the image cube we glued the two subbands together by
using the AIPS task UJOIN, the final dataset has a bandwidth of $\sim52$~MHz ($\sim650$~\kms) centered at the
velocity 7327.0~\kms. Similarly to the VLBA data reduction, before producing the image cubes we subtracted the
continuum in the uv-plane using the AIPS task UVLSF. We unsuccessfully tried to self-calibrate the brightest maser
channels. Also in this case we searched for maser features by using the process described in \citet{sur11}
(detection threshold $>3\sigma$) and we determined the absolute position by fitting the brightest maser spot of
each maser feature with the AIPS task JMFIT. Also in this case we considered conservative position errors
equal to half the beam projected along $\alpha_{2000}$ and $\delta_{2000}$, which are $\pm0.5$~mas and $\pm1.9$~mas,
respectively.\\
\indent We averaged the channels without \water ~maser emission to produce the continuum map of the nuclear region
of \txs.
\subsection{\object{New observations: EVN data epoch 2018.44}}
\label{evn_obspol}
To attempt for the first time to detect polarized emission from an extra-galactic \water ~maser, we performed a second
EVN epoch in full polarization mode. This epoch was observed with seven of the EVN antennas on June 9, 2018. 
To avoid the DBBC firmware issue that affected epoch 2017.83, we used four subbands of 16 MHz each ($\sim217$~\kms) with 
an overlap between them of 5~MHz. Therefore, we covered a total velocity range of $\sim660$~\kms. Because the main 
objective of these observations was to detect polarized \water ~maser emission, for which typical linear and circular 
polarization fractions in Galactic maser are of few percent (e.g., \citealt{sur11}), to maximize our on-source time we did not 
observe the phase-reference calibrator as done in epoch 2017.83. Indeed, we increased the observation time of \txs ~up to
about 4 h, i.e., one hour more than in epoch 2017.83, and we observed every hour 
the polarization calibrator J2202+4216. By including the overheads, the total observation time was 6 h. The data were 
also in this case correlated with the SFXC at the JIVE, but this time generating all four polarization combinations 
(RR, LL, RL, LR) by using 4096 channels per subband (channel width $\sim~4$~kHz$\sim~0.05$~\kms). The narrow channel width
is necessary in order to properly measure the Zeeman-splitting from the circularly polarized spectra of the maser features
in case circular polarization is detected. The observational details are reported in Table~\ref{Obs}.\\
\indent The data were calibrated using AIPS by following the procedure described in \citet{sur11}, i.e., after applying the
a-priori calibration obtained from the JIVE pipeline, we performed the bandpass, delay, phase, and polarization calibration
on the polarization calibrator J2202+4216. After applying the solutions to \txs, first we shifted the frequency using the 
AIPS task CVEL and we glued the four subbands together similarly as we did in epoch 2017.83, then we successfully 
self-calibrated 90 spectral channels ($\sim350$~kHz~$\sim4.5$~\kms) around the peak of the brightest maser feature 
(7398.9~\kms). Finally we imaged the \textit{I}, \textit{Q}, \textit{U}, and \textit{V} Stokes cubes ($\rm{rms}\approx
5$~\mjyb). The \textit{Q} and \textit{U} cubes were combined using the AIPS task COMB to produce cubes of linearly
polarized intensity ($\rm{POLI}=\sqrt{Q^2+U^2}$) and polarization angle ($\chi=\frac{1}{2}~\arctan(\frac{U}{Q})$). The
searching for maser features was performed using the process described in \citet{sur11} (detection threshold $>3\sigma$).\\
\indent Also for this EVN epoch we averaged the channels without \water ~maser emission to produce the continuum map of
the nuclear region of \txs.
\section{Results}
\label{res}
Here, we report separately the results obtained from the four VLBI epochs in chronological order.
\begin{figure*}[th!]
\centering
\includegraphics[width = 6 cm]{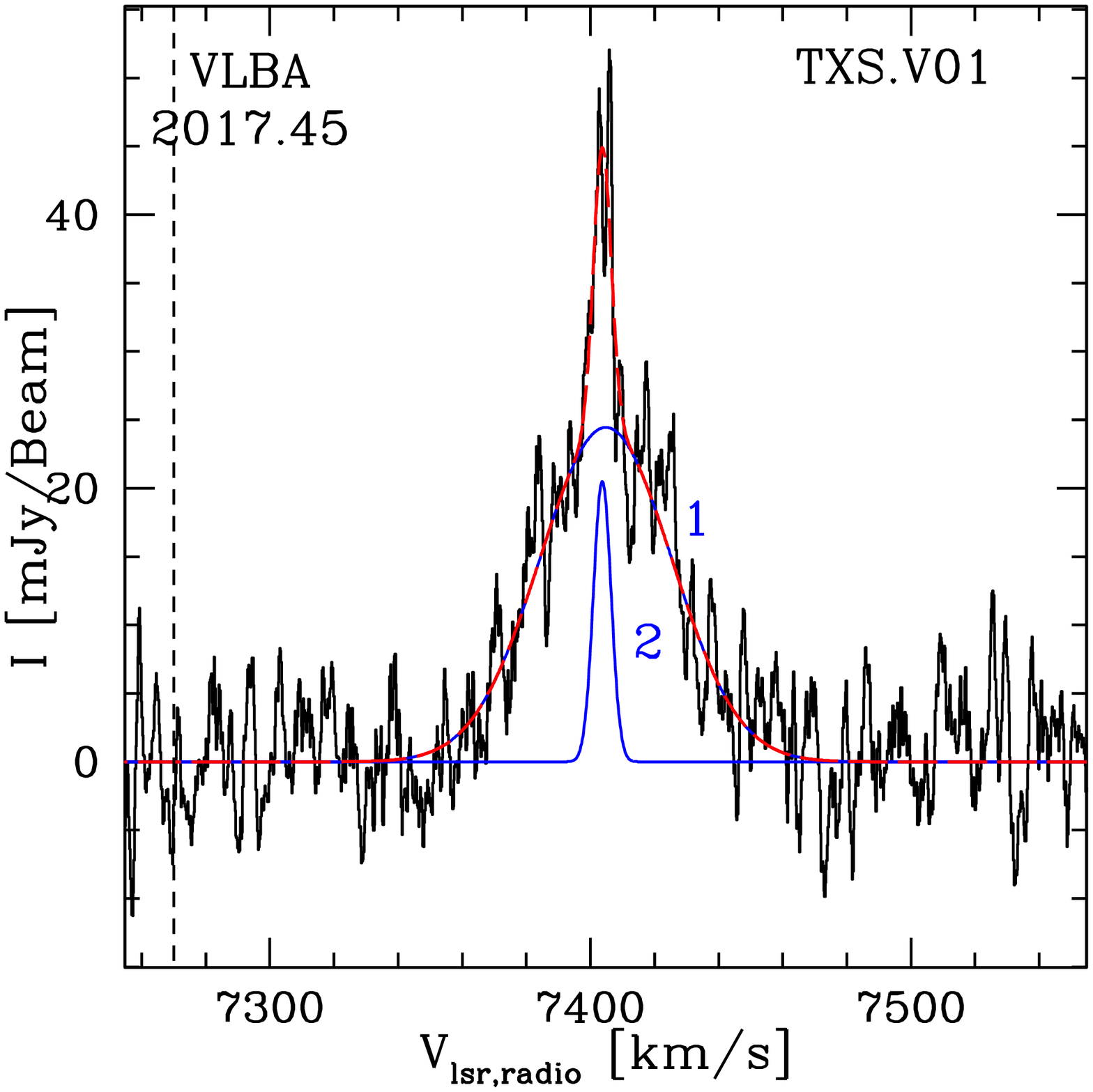}
\includegraphics[width = 6 cm]{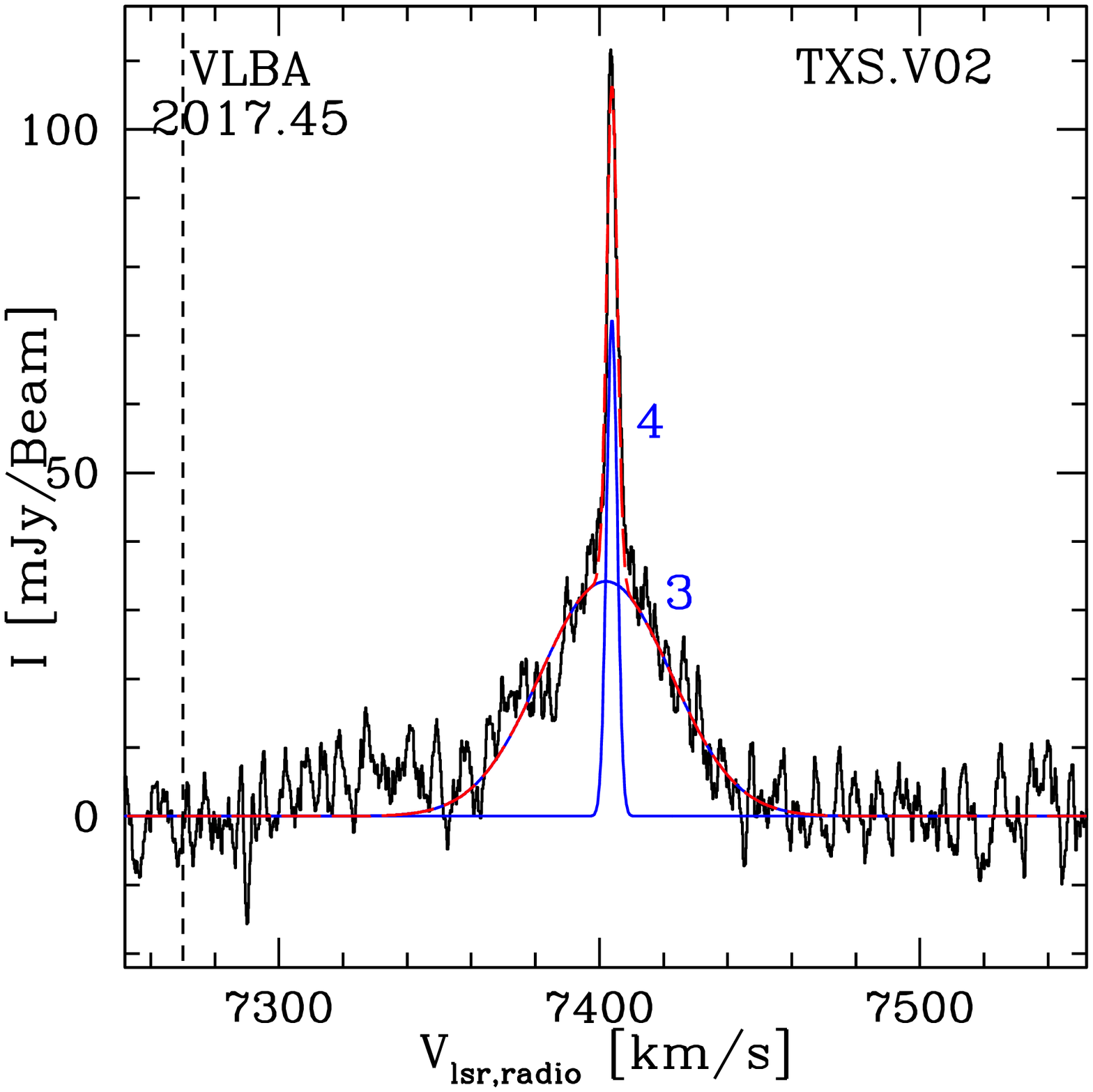}
\includegraphics[width = 6 cm]{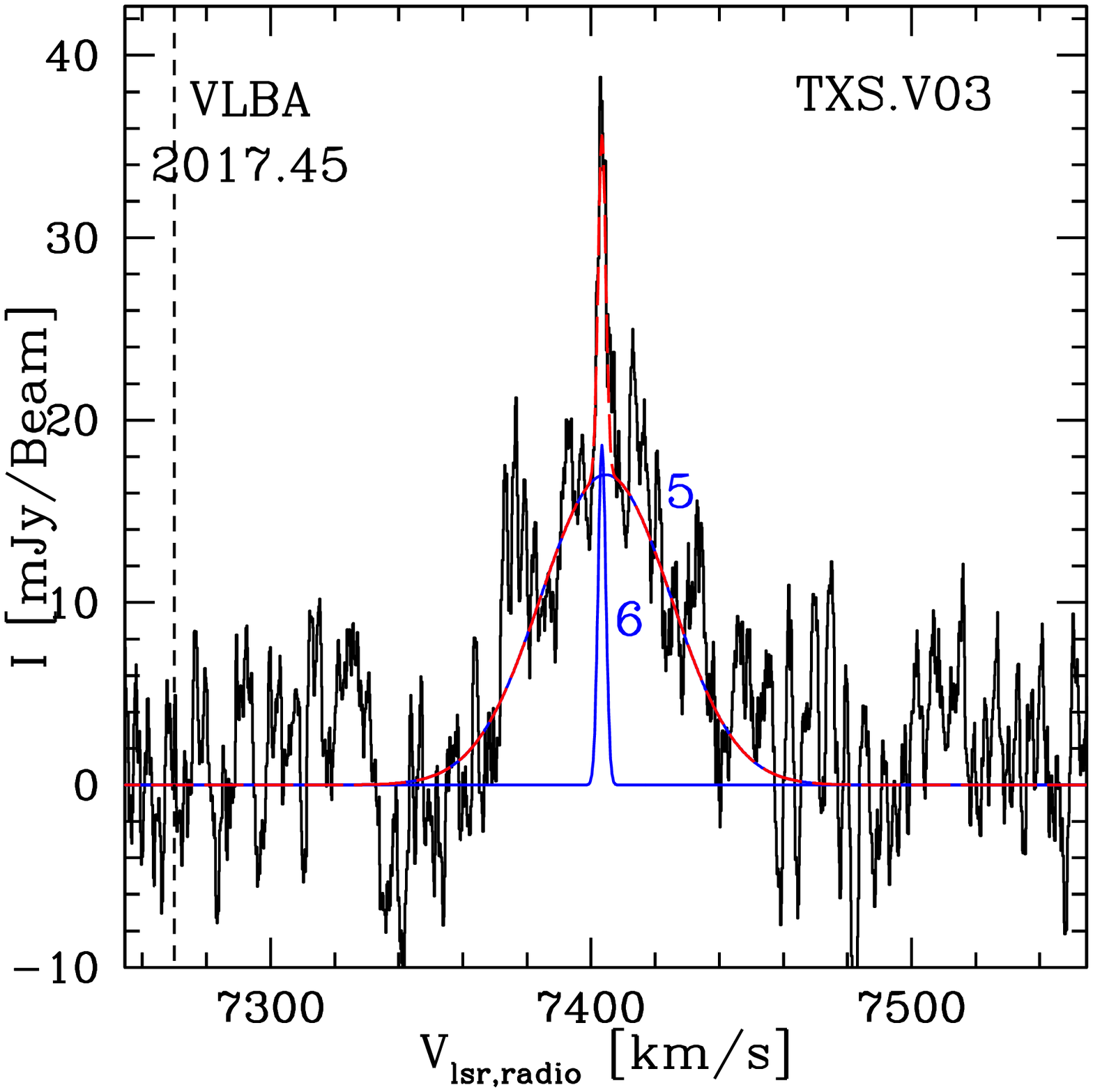}
\includegraphics[width = 6 cm]{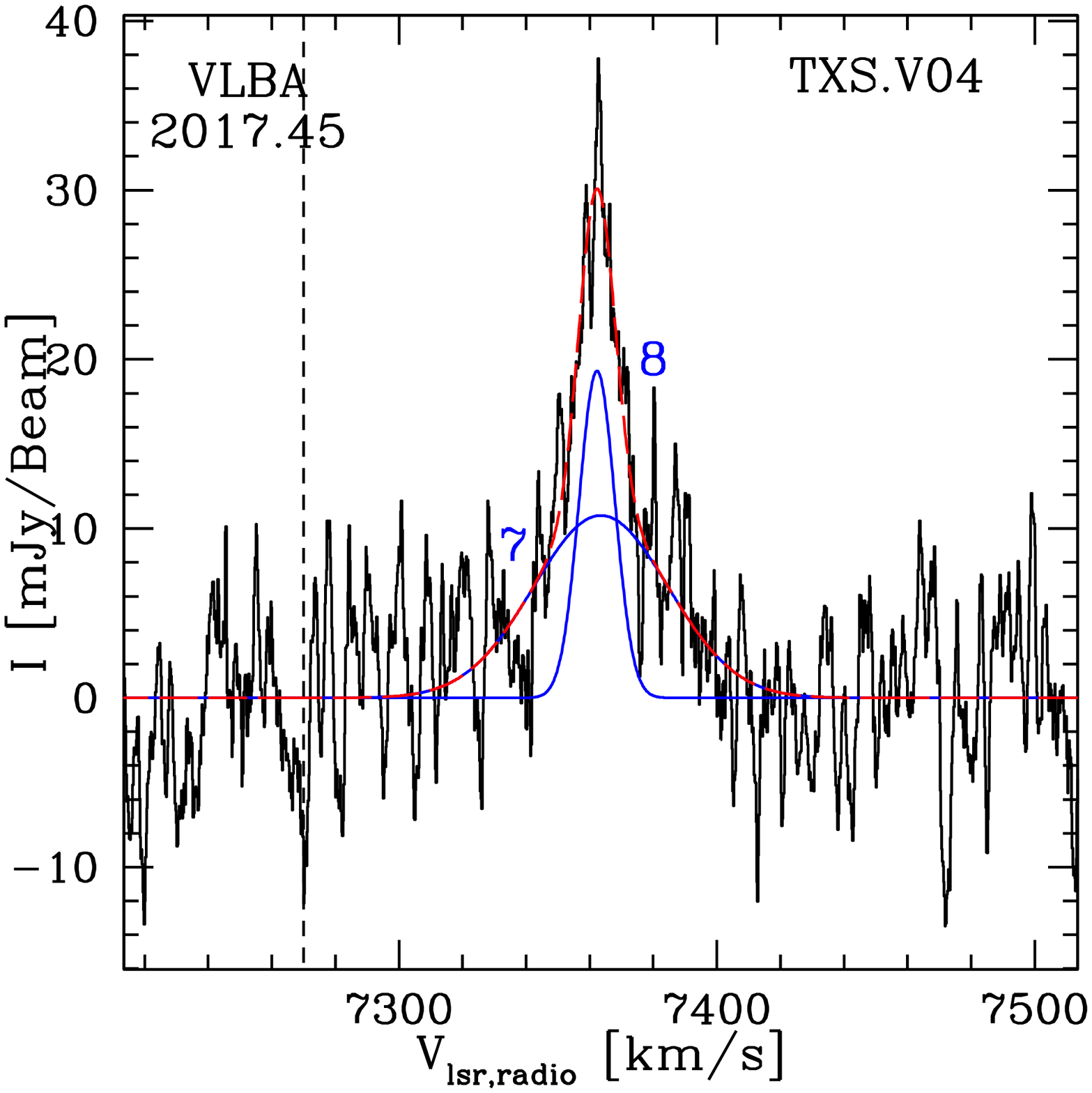}
\includegraphics[width = 6 cm]{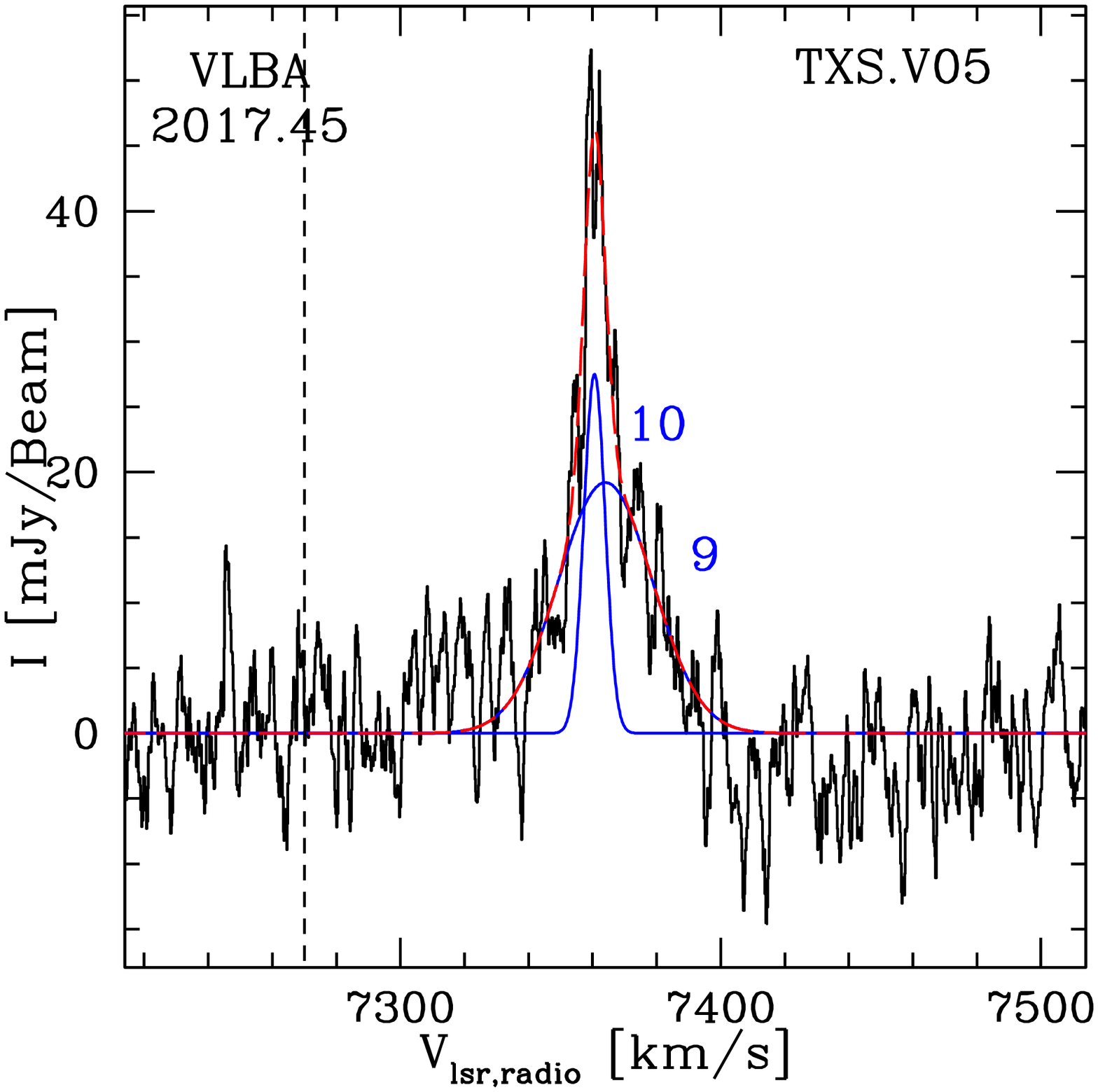}
\includegraphics[width = 6 cm]{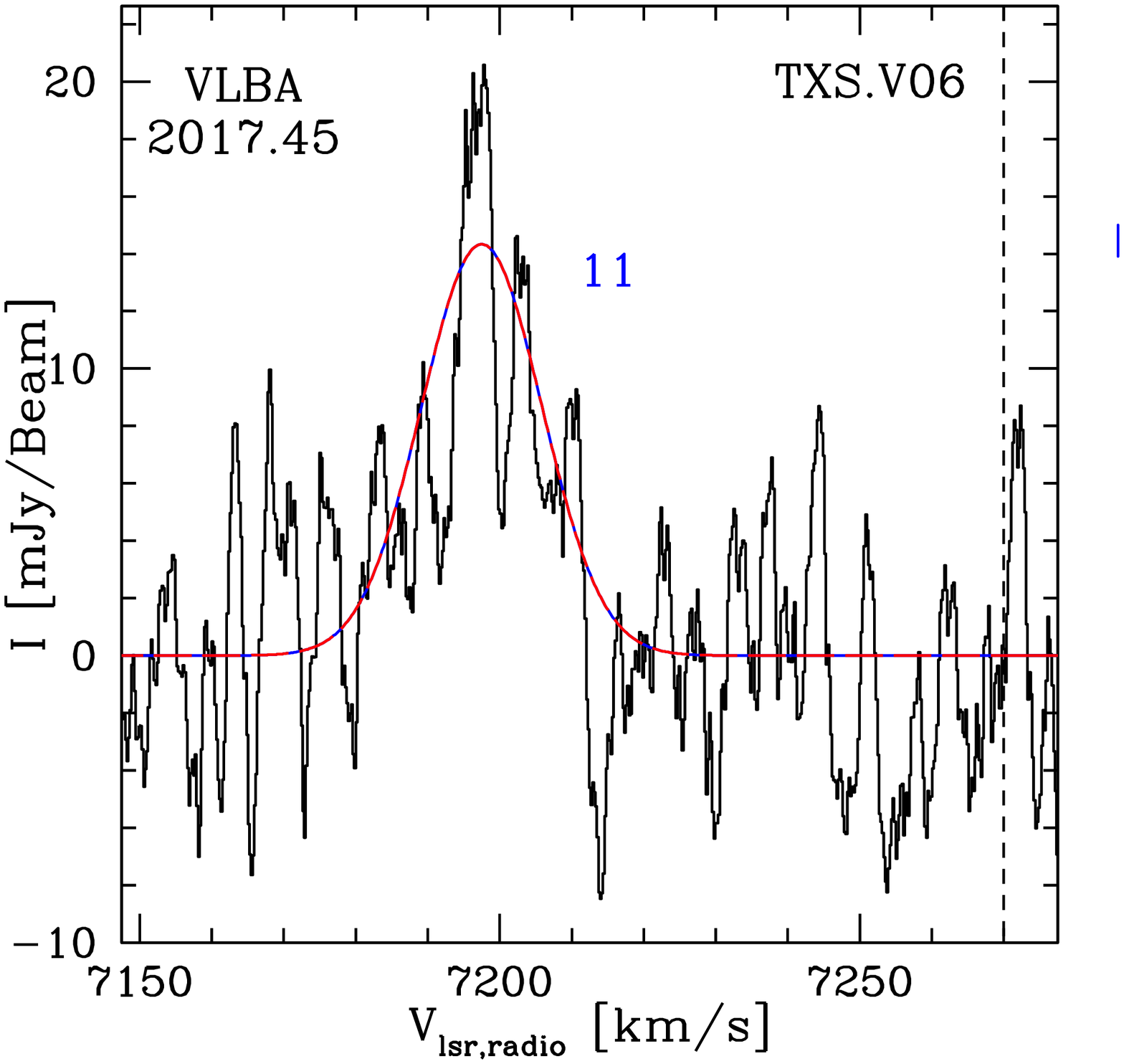}
\caption{Total intensity spectra ($I$) of the \water ~maser features detected towards \txs ~with the VLBA in epoch
2017.45 after a boxcar smoothing of 10 channels. The thick blue lines are the Gaussian fit of the different 
components that are labeled in blue (see Table~\ref{2017_tab}), the dashed red line is the combination of the
components for each maser feature. The dashed black line indicates the systemic velocity of the galaxy in the radio convention (i.e., 7270~\kms).}.
\label{vlba_2017}
\end{figure*}
\begin{figure*}[h!]
\centering
\includegraphics[width = 6 cm]{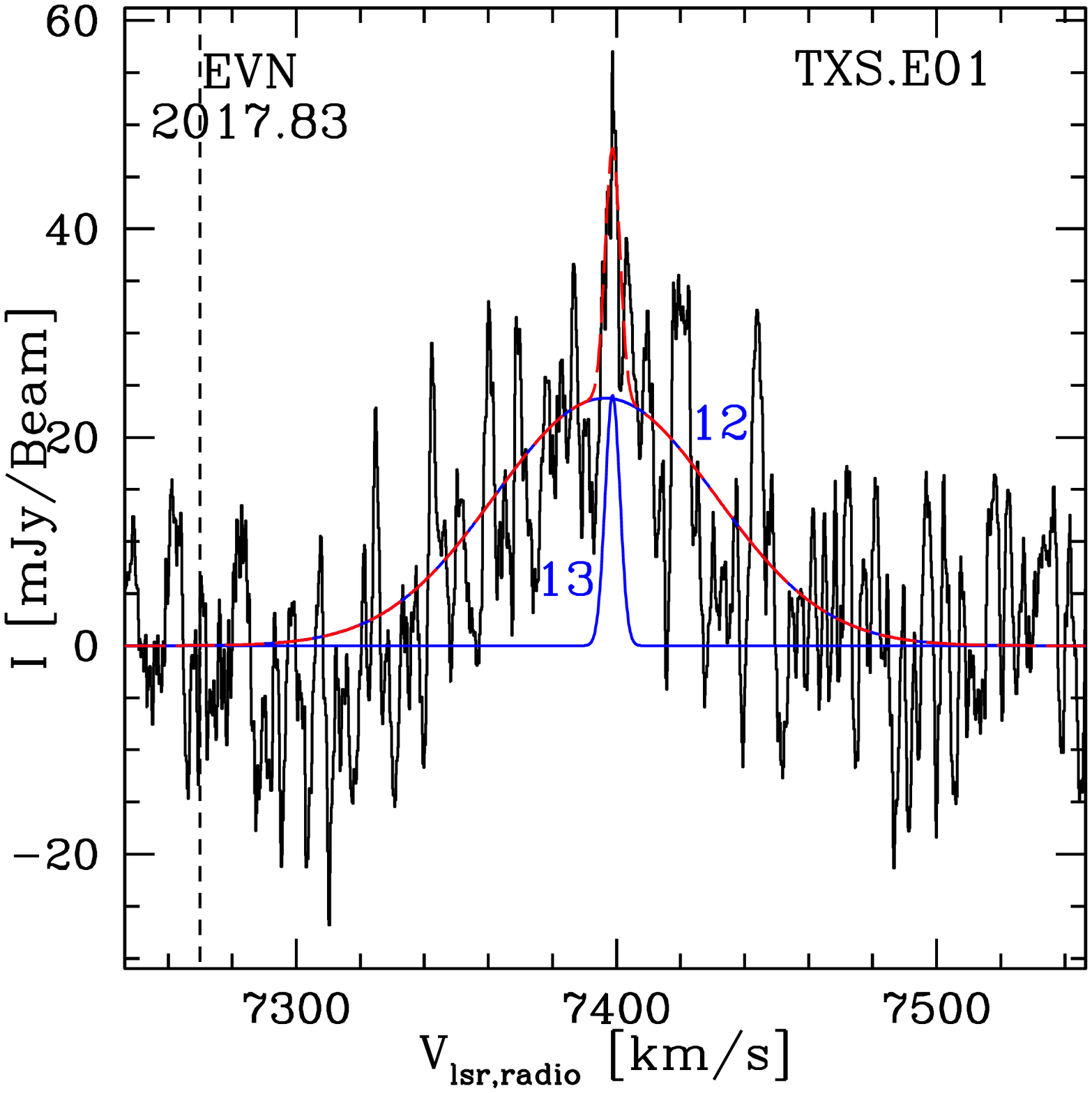}
\includegraphics[width = 6 cm]{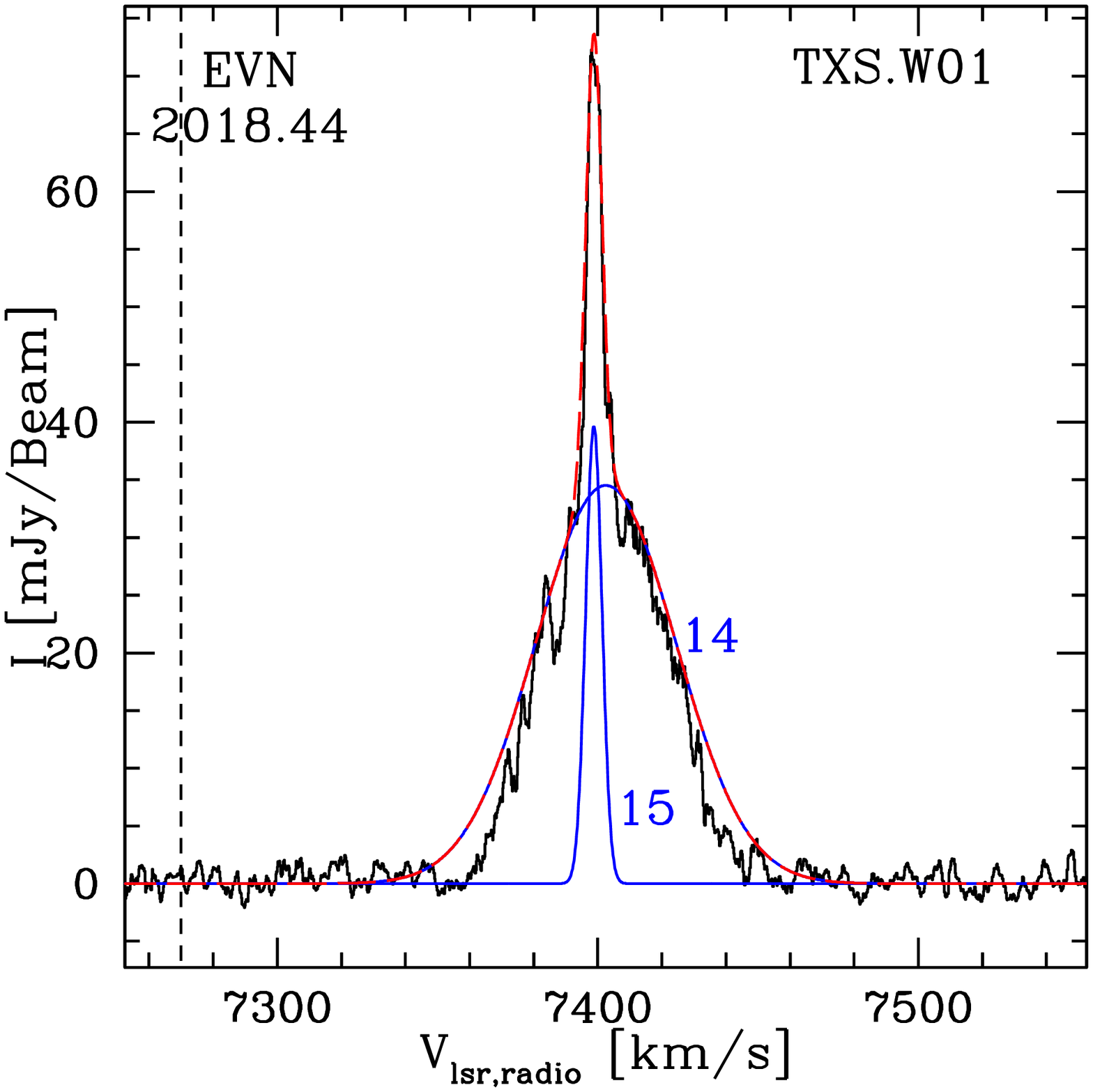}
\includegraphics[width = 6 cm]{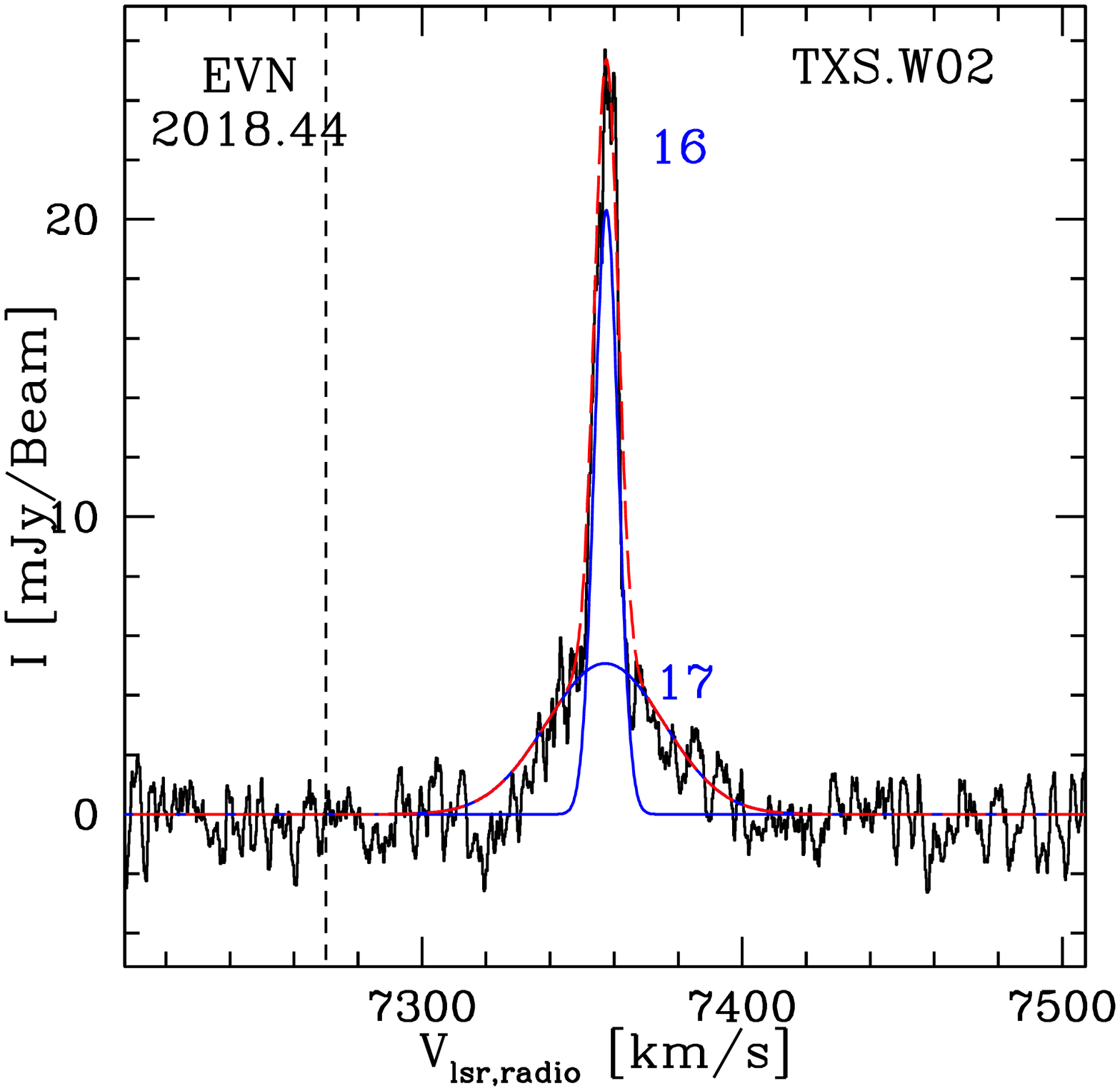}
\caption{Total intensity spectra ($I$) of the \water ~maser features detected towards \txs ~in epochs 2017.83 (first
panel on the left) and 2018.44 (second and third panels) with the EVN after a boxcar smoothing of 10 channels. A
channel averaging has been performed in epoch 2018.44 in order to have comparable spectral width between the two EVN
epochs. The thick blue lines are the Gaussian fit of the different components labeled in blue (see
Table~\ref{2017_tab}), the dashed red line is the combination of the components fit for each maser
feature. The dashed black line indicates the systemic velocity of the galaxy in the radio convention (i.e., 7270~\kms).}.
\label{txs_evn}
\end{figure*}
\begin {table*}[ht]
\caption []{Parameters of the 22 GHz \water ~maser features detected towards \txs ~in epochs 2017.45 (VLBA), 2017.83
and 2018.44 (EVN). The columns are: the maser feature name, absolute position (right ascension and declination) or
offset position with respect to the reference maser feature (for epoch 2018.44), identified component from the 
Gaussian fit, peak flux density, peak velocity (radio convention), linewidth of the maser feature, spectral noise per 
channel, signal-to-noise ratio, spectral noise per \kms, integrated flux, isotropic luminosity. } 
\begin{center}
\scriptsize
\begin{tabular}{ l c c c c c c c c c c c}
\hline
\hline
\,\,\,\,\,(1)&(2)                     & (3)                                     & (4)       & (5)        & (6)           &(7)                                   &(8)                       &(9)  & (10) & (11) & (12) \\
Maser     &  $\alpha_{2000}$\tablefootmark{a} & $\delta_{2000}$\tablefootmark{b}& Comp. & Peak flux\tablefootmark{c}  & $V_{\rm{lsr, radio}}$\tablefootmark{c}& $\Delta v\rm{_{L}}$\tablefootmark{c} & rms\tablefootmark{c,d}     & SNR & rms & $ \int S dV$\tablefootmark{c} & $L_{\rm{H_2O}}$\tablefootmark{e}\\
          &                           &                                         &           &  Density(I) &               &                     &         &   & &\\
          & ($\rm{^{h}:~^{m}:~^{s}}$) & ($\rm{^{\circ}:\,':\,''}$)              &           & (\mjyb)    &      (\kms)   &      (\kms)         & (\mjybc) & & (\mjybks)& (Jy \kms) &($L_{\odot}$)\\ 
          \\
\hline
\multicolumn{11}{c}{\textbf{VLBA}}\\
\hline
TXS.V01   & 22:29:12.4943            &  -18:10:47.2369         &  1   & $24.45$ &  $7404.8\pm0.4$  &  $48.5\pm0.9$ & 4.0 &  11\tablefootmark{f}& 1.83 & $1.26\pm0.02$ & $369$\tablefootmark{f}\\
          &                          &                         &  2   & $20.50$ &  $7403.7\pm0.2$  &  $6.5\pm0.3$  &     &      &                 & $0.14\pm0.01$ & \\
TXS.V02   & 22:29:12.4943            &  -18:10:47.2367         &  3   & $34.19$ &  $7402.0\pm0.3$  &  $47.9\pm0.9$ & 4.7 &  23\tablefootmark{f} & 2.15 & $1.74\pm0.02$ & $537$\tablefootmark{f}    \\
          &                          &                         &  4   & $72.13$ &  $7403.91\pm0.04$&  $3.9\pm0.1$  &     &      &                & $0.30\pm0.01$ & \\ 
TXS.V03   & 22:29:12.4944            &  -18:10:47.2359         &  5   & $17.01$ &  $7404.6\pm0.6$  &  $47.8\pm1.3$ & 4.5 & 8\tablefootmark{f}  & 2.06 & $0.86\pm0.02$ & $242$\tablefootmark{f}\\ 
          &                          &                         &  6   & $18.62$ &  $7403.4\pm0.1$  &  $2.9\pm0.3$  &     &      &                & $0.06\pm0.01$ & \\ 
TXS.V04   & 22:29:12.4944            &  -18:10:47.2338         &  7   & $10.78$ &  $7363.5\pm1.1$  &  $50.0\pm3.8$ & 4.6 & 6\tablefootmark{f}  & 2.11 & $0.57\pm0.04$ & $221$\tablefootmark{f}   \\ 
          &                          &                         &  8   & $19.32$ &  $7362.3\pm0.3$  &  $13.2\pm1.1$ &     &      &                & $0.27\pm0.03$ & \\ 
TXS.V05   & 22:29:12.4945            &  -18:10:47.2334         &  9   & $19.21$ &  $7364.0\pm0.5$  &  $34.7\pm1.4$ & 4.4 & 11\tablefootmark{f} & 2.02 & $0.71\pm0.23$ & $248$\tablefootmark{f} \\ 
          &                          &                         &  10  & $27.51$ &  $7360.6\pm0.2$  &  $7.8\pm0.5$  &     &      &                & $0.23\pm0.02$ &  \\ 
TXS.V06   & 22:29:12.4942            & -18:10:47.2421          &  11  & $14.33$ &  $7197.5\pm0.4$  &  $19.7\pm1.3$ & 4.3 & 3  & 1.97                 & $0.30\pm0.01$ & $79$ \\
\hline
\multicolumn{11}{c}{\textbf{EVN}}\\
\hline
\multicolumn{11}{c}{2017.83}\\
\hline
TXS.E01   & 22:29:12.4944            &   -18:10:47.2367        &  12  & $23.76$ &  $7396.5\pm1.0$  & $68.9\pm2.4$  & 9.5 & 5\tablefootmark{f}  & 4.35 & $1.74\pm0.05$ & $491$\tablefootmark{f}\\
          &                &          &                         &  13  & $24.03$ &  $7398.8\pm0.3$  & $4.6\pm1.1$   &     &                       & $0.12\pm0.02$ &\\ 
\hline
\multicolumn{11}{c}{2018.44}\\
\hline
Maser     &  $\alpha_{2000}$  & $\delta_{2000}$ & Component & Peak flux\tablefootmark{h}  & $V_{\rm{lsr, radio}}$\tablefootmark{h}& $\Delta v\rm{_{L}}$\tablefootmark{h} & rms\tablefootmark{h,d}     & SNR & rms & $ \int S dV$\tablefootmark{h} & $L_{\rm{H_2O}}$\tablefootmark{e}\\
          & offset\tablefootmark{g} & offset\tablefootmark{g} &           &  Density(I) &               &                     &         &  & & &\\
          &      (mas)        &     (mas)       &           & (\mjyb)     &      (\kms)   &      (\kms)         & (\mjybc) & &(\mjybks)& (Jy \kms) &($L_{\odot}$)\\ 
          \\
\hline
TXS.W01   & 0.000             &   0.000         & 14        & $34.52$     & $7402.5\pm0.1$& $43.7\pm0.1$   & 2.0 & 24\tablefootmark{f} & 0.92 & $1.605\pm0.005$ & $480$\tablefootmark{f}\\
          &                   &                 & 15        & $39.60$     & $7398.8\pm0.1$& $5.01\pm0.1$   &     &                   &   & $0.215\pm0.002$ &\\ 
TXS.W02   & +2.608          &  +3.998         & 16        & $5.07$      & $7357.2\pm0.4$& $36.6\pm1.0$   & 1.3 & 34\tablefootmark{f} & 0.60 & $0.198\pm0.005$ & $94$\tablefootmark{f}\\
          &                   &                 & 17        & $20.31$     & $7357.6\pm0.1$& $7.4\pm0.1$    &     &                   &    & $0.160\pm0.003$ & \\
\hline
\end{tabular} \end{center}
\tablefoot{
\tablefoottext{a}{VLBA: the error is $\pm0.2$~mas. EVN (epoch 2017.83): the error is $\pm0.5$~mas.}
\tablefoottext{b}{VLBA: the error is $\pm0.7$~mas. EVN (epoch 2017.83): the error is $\pm1.9$~mas.}
\tablefoottext{c}{Gaussian fit results after a boxcar smoothing of 10 channels.}
\tablefoottext{d}{The channel width is 0.21~\kms.}
\tablefoottext{e}{Isotropic luminosities are derived from $L_{\rm{H_{2}O}}/[\solum]=0.023 \times \int S \, \rm{d}\textit{V}/[\rm{Jy \ km \ s^{-1}}] \times \textit{D}^{2}/[\rm{Mpc^{2}}]$, where  $D=107.1$~Mpc (derived assuming $H_{\rm{0}}=70$~\kmsM, \citealt{kuo18}).}
\tablefoottext{f}{Estimated considering the sum of the two components.}
\tablefoottext{g}{No absolute position has been measured in this epoch.}
\tablefoottext{h}{Gaussian fit results after channels averaging to have a spectral channel width identical to
epoch 2017.83 and after a boxcar smoothing of 10 channels.}
}
\label{2017_tab}
\end{table*}
\subsection{\object{VLBA epoch 1998.40}}
\label{1998_res}
We identified three (red-shifted with respect to the systemic velocity) out of the seven \water ~maser features 
reported by \citet{bal05}. We named them as
TXS.B01--TXS.B03 and they are shown in Fig.~\ref{vlba_1998} (left panel) and their parameters are reported in
Table~\ref{1998_tab}. Blue-shifted maser features did not emerge at $3\sigma$ in our image cube neither at the 
velocity reported by \citet{bal05}, i.e, $V_{\rm{lsr,radio}}=7100$~\kms ~and $V_{\rm{lsr,radio}}=7200$~\kms, nor at the
positions indicated by \citet{bal05} (blue circles in Fig.~\ref{posplot_vlba}). However, note that the velocities reported
by \citet{bal05} are at the edges of the subbands where the response of the backend is not linear and hence more sensitive to
small differences in data reduction and/or analysis. The \water ~maser features are linearly
distributed from northeast (the least red-shifted and luminous feature) to southwest (the most red-shifted and
luminous feature) with a position angle of $\rm{PA}=+26$\d$\pm16$\d ~(see Fig.~\ref{posplot_vlba}). 
No position-velocity gradients are observed. From a multiple
Gaussian fit (see Fig.~\ref{vlba_1998}) we determined that the most luminous maser features (TXS.B01 and TXS.B02)
are composted of two broad Gaussian components (see Table~\ref{1998_tab}) with respect to typical individual maser
features associated with AGN disks or star-formation sites. One component is much broader than the other one and all
components have linewidth ($\Delta v\rm{_{L}}$) greater than 20~\kms, including TXS.B03. We note that the Gaussian 
component number 4 of feature TXS.B02 shows a $\Delta v\rm{_{L}}$ double than the second broadest one (component number 
1 of TXS.B01; see Table~\ref{1998_tab}). The least luminous \water ~maser feature TXS.B03 ($L_{\rm{H_2O}}=366~L_{\odot}$) 
is instead composed of only one relatively narrow Gaussian component ($\Delta v\rm{_{L}}\approx26$~\kms). No continuum
emission has been detected above a $1\sigma$ noise level of $0.5$~\mjyb ~(in the map produced averaging the line-free channels).
\subsection{\object{VLBA epoch 2017.45}}
\label{2017_res}
We detected six \water ~maser features in epoch 2017.45, their profiles are plotted in Fig.~\ref{vlba_2017} and 
their linear distribution ($PA=+28$\d$\pm12$\d) is shown in Fig.~\ref{posplot_vlba} (right panel). The maser
features are listed in Table~\ref{2017_tab} where we report the absolute position, the results of the multiple
Gaussian fit, and the isotropic luminosity. The most red-shifted maser features are located at the center of the
distribution (TXS.V01--TXS.V03), where the most luminous maser feature TXS.V02 is detected
($L_{\rm{H_2O}}=537~L_{\odot}$). The other (less) red-shifted maser features (TXS.V05 and TXS.V04) are located
northeast and the only blue-shifted maser feature (TXS.V06), which is the least luminous
($L_{\rm{H_2O}}=79~L_{\odot}$), is located southwest. Also in this case no position-velocity gradients are observed. 
From the multiple Gaussian fit we find that all the
red-shifted maser features are composed of two Gaussian components, one of which is very narrow 
($2.9$~\kms$\leq\Delta v\rm{_{L}}\leq13.2$~\kms) while the other one is broad 
($34.7$~\kms$\leq\Delta v\rm{_{L}}\leq50$~\kms; see Table~\ref{2017_tab} and Fig.~\ref{vlba_2017}). TXS.V06 is 
the only one composed of one Gaussian component. No continuum emission has been detected at $1\sigma$ level 
($\rm{rms}=0.4$~\mjyb).
\subsection{\object{EVN epoch 2017.83}}
\label{2017_res_evn}
One \water ~maser feature (named TXS.E01) has been detected with the EVN in epoch 2017.83. This is listed in
Table~\ref{2017_tab} with its spectrum showed in Fig.~\ref{txs_evn} (left panel). TXS.E01 is red-shifted with respect
to the systemic velocity of the galaxy and its absolute position is $\alpha_{2000}=22^{h}29^{m}12.4944^{s}$ and
$\delta_{2000}=-18^{\circ}10'47.\!\!''2367$. This maser feature does not show any position-velocity gradient.
The multiple Gaussian fit of its spectrum reveals that the maser
feature ($L_{\rm{H_2O}}=491~L_{\odot}$) is composed of two Gaussian components with similar peak flux densities
($\sim24$~\mjyb). The narrowest one ($\Delta v\rm{_{L}}=4.6$~\kms) is centered at 7398.8~\kms, and the other one
($\Delta v\rm{_{L}}=68.9$~\kms) at 7396.5~\kms. No continuum emission has been detected at $1\sigma$ level
($\rm{rms}=0.2$~\mjyb).
\begin{figure}[t!]
\centering
\includegraphics[width = 9 cm]{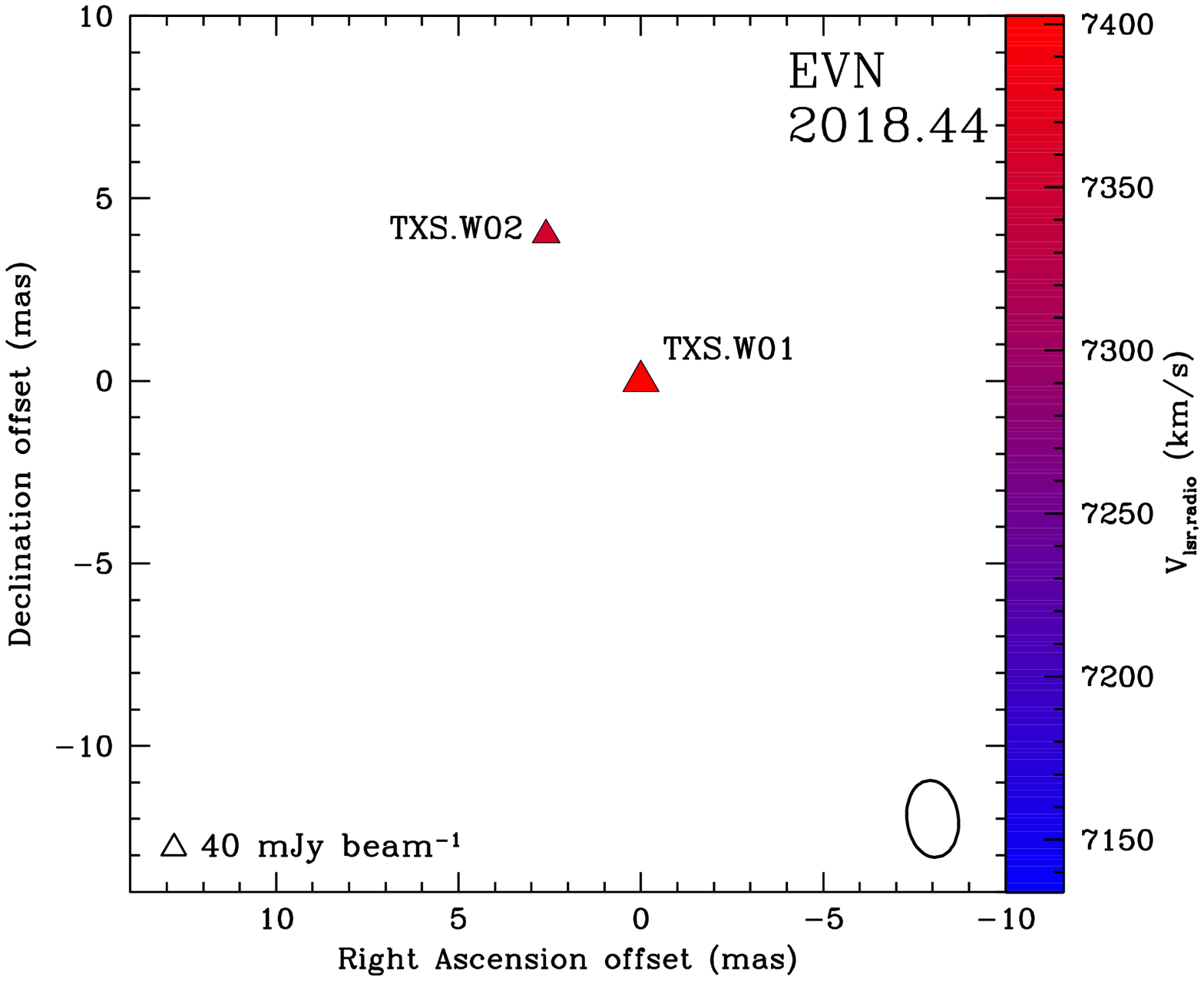}
\caption{View of the \water ~maser features detected towards \txs ~with the EVN in epoch 2018.44. Same symbols 
as in Fig.~\ref{posplot_vlba}. In absence of absolute position measurements the relative positions of all maser features are
evaluated considering the brightest maser feature (TXS.W01) as reference.}
\label{posplot_evn}
\end{figure}
\subsection{\object{EVN epoch 2018.44}}
In epoch 2018.44 we detected two \water ~maser features (named TXS.W01 and TXS.W02) that are listed in Table~\ref{2017_tab},
here instead of the absolute position, we report the relative position with respect to the maser feature used to
self-calibrate the data, i.e. TXS.W01. Both maser features do not show a position-velocity gradient. Their spectra are shown 
in Fig.~\ref{txs_evn}. Both maser features are red-shifted
with respect to the systemic velocity of the galaxy and are aligned from northeast (7357.6~\kms) to southwest (most
red-shifted; i.e. 7402.5~\kms). Again the maser spectra are better fit by two Gaussian components, one narrow
($5.0$~\kms$\leq\Delta
v\rm{_{L}}\leq7.4$~\kms) and one wide ($36.6$~\kms$\leq\Delta v\rm{_{L}}\leq43.7$~\kms). In the case of TXS.W01 
($L_{\rm{H_2O}}=480~L_{\odot}$) the peak flux of components 14 and 15 are close, while for TXS.W02
($L_{\rm{H_2O}}=94~L_{\odot}$) we have a large difference between the two components 16 and 17. None of the maser features
showed a signature of polarized emission, neither linearly nor circularly polarized ($P_{\rm{l,V}}^{\rm{W01}}<8\%$ and
$P_{\rm{l,V}}^{\rm{W02}}<15\%$). No continuum emission has been detected at $1\sigma$ level ($\rm{rms}=0.2$~\mjyb).
\section{Discussion}
\label{discussion}
\begin{figure}[t!]
\centering
\includegraphics[width = 8.5 cm]{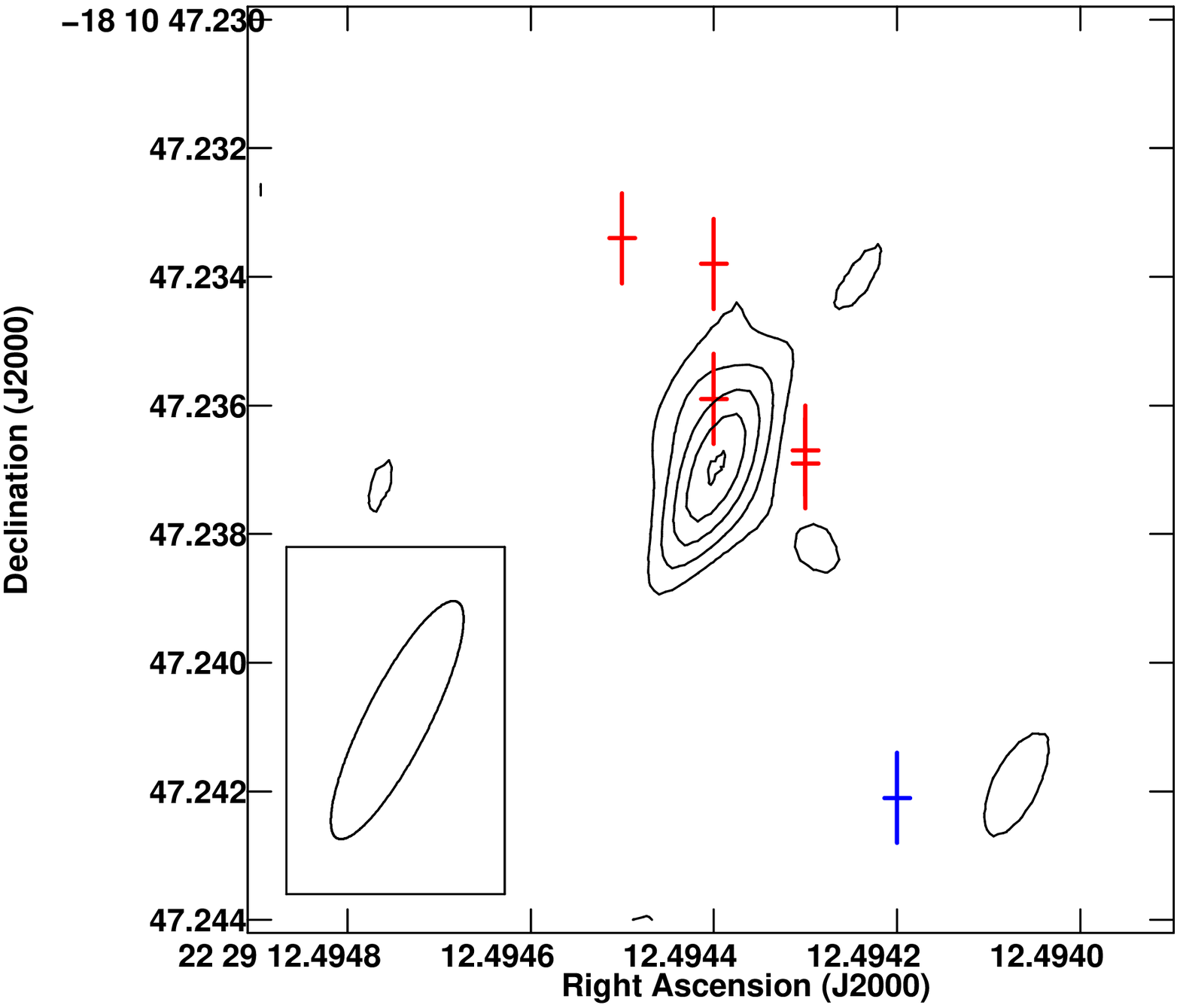}
\caption{Comparison of the absolute positions measured for TXS.E01 (contour plot of the peak channel) and
for TXS.V01--TXS.V06 (crosses). Contours are $2, 3, 4, 5, 6 \times 21.75$~\mjyb ~and on the bottom left
corner the EVN beam size is shown (see Table~\ref{Obs}). Red crosses indicate the red-shifted \water ~maser features with
respect to the systemic velocity of the galaxy ($V_{\rm{lsr,radio}}^{\rm{TXS}}=7270$~\kms), and the blue cross the blue-shifted
one. The size of the crosses correspond to the errors of the absolute positions.}
\label{vlba_evn}
\end{figure}
In the following, we compare the different epochs and we discuss the possible origin of the
22~GHz \water ~maser emission in light of the new findings. Regarding the VLBA epoch~1998.40, we will discuss the results
obtained by our calibration and analysis and only marginally we refer to those reported in \citet{bal05}.
\subsection{Comparison between epochs.}
\label{comp}
Before comparing the results obtained from epoch 1998.40 with those from the most recent epochs, we quickly discuss the
contribution provided by the EVN epochs 2017.83 and 2018.44.\\
\indent The only maser feature (TXS.E01) detected during the EVN observations in epoch 2017.83 does not add significant
information to the maser distribution than those we already have from the VLBA epoch 2017.45. This is mainly due to the
worse spatial resolution (the beam is almost three times larger) and to the higher spectral noise (see Table~\ref{Obs}).
Furthermore, the amplitude uncertainty due to the technical problem that affected the EVN observations in epoch 2017.83 
prevent the possibility to study the flux variation at time scale of months between the three most recent epochs. However,
the results of this EVN epoch still play an important part in our discussion. Indeed, although many of the \water ~maser
features detected with the VLBA in epoch 2017.45 are undetected with the EVN in epoch 2017.83 due to the high noise that
hides them below our detection threshold, the registered absolute position of TXS.E01 agrees with those measured for
the maser features detected with the VLBA. This consistency is evident in Fig.~\ref{vlba_evn}, where we have overlapped the
absolute positions of the \water ~maser features detected with the VLBA in epoch 2017.45 (crosses) to the contours map of the
brightest maser spot of TXS.E01. \\
The maser features detected during the EVN epoch 2018.44 confirm the presence of the persistent linear
distribution of the maser features over time. In particular, we can state that TXS.W01 is produced by the blending of
features TXS.V01--V03, by comparing $\Delta v\rm{_{L}}$ and the velocity range, and that TXS.W02 is related to TXS.V05 
although their relative position with respect to the corresponding brightest maser features is a bit different, in epoch 
2018.44 it is larger. Also note that the blending of maser features is expected because of the beam shape of epoch 2018.44 
that is elongated towards northeast.
Furthermore, we notice a small velocity difference between TXS.W02 and TXS.V05. However, both these might be due to the small
difference in beam size between the two epochs. It is more important instead that the profile morphology of the maser
features detected with the EVN (Fig.~\ref{txs_evn}), even though these are the result of a blending of several maser features
detected during the VLBA observations, confirms those observed for TXS.V01--TXS.V05 (see Fig.~\ref{vlba_2017}). All spectra 
are composed of two Gaussian components, one much narrower than the other, as observed for the red-shifted maser features
with the VLBA in epoch 2017.45. Of course the last EVN epoch also indicate the absence of a strong magnetic field able to
produce a circularly polarized emission. Therefore, the EVN results from both epochs firmly confirm the results
that we obtained with the VLBA in epoch 2017.45. From now on our discussion will be focused only on the two VLBA epochs
(i.e., 1998.40 and 2017.45).\\
\begin{figure}[th!]
\centering
\includegraphics[width = 9 cm]{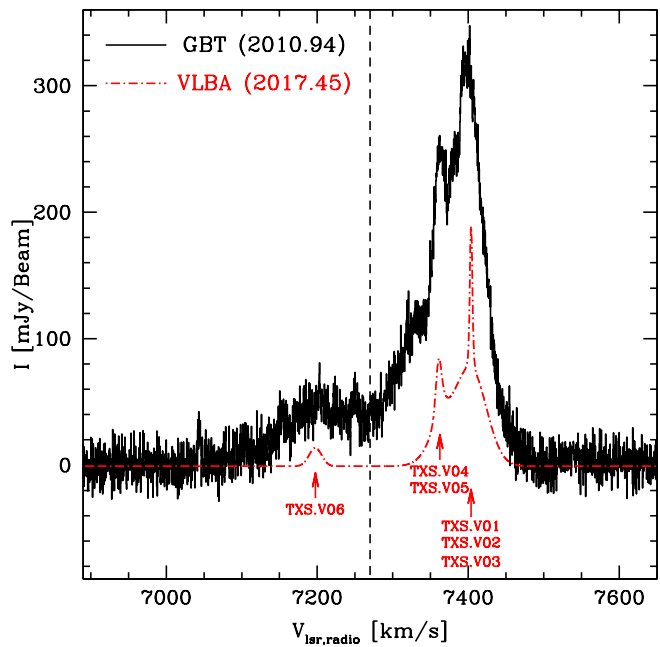}
\caption{Comparison between the total intensity spectra ($I$) of the \water ~gigamaser detected towards \txs
~with the GBT in epoch 2010.94 (black histogram) and the combination (red dot-dashed line) of all the eleven
Gaussian components determined for the six \water ~maser features (TXS.V01--TXS.V06) detected with the VLBA in epoch
2017.45. The red arrows indicate the velocities of the total intensity peak of the different maser features (see
Table~\ref{2017_tab}). The dashed black line indicates the systemic velocity of the galaxy in the radio convention (i.e.,
7270~\kms).}
\label{GBT}
\end{figure}
\indent We summed the eleven Gaussian components obtained by fitting the six maser features detected with
the VLBA in epoch 2017.45, the resultant profile is showed as red dot-dashed line in Fig.~\ref{GBT}. Here, we also show the
spectrum of the \water ~gigamaser detected with the GBT in epoch 2010.94. We can see that the two profiles are consistent with
each other, even though the fluxes are different. This might be due to the different spatial resolution of the two observations and, more
importantly, on the criteria we chose for identifying the \water ~maser features in the VLBA data cube and the way the spectra 
are extracted from the VLBI image cube (see Sect.~\ref{vlba_obs}). The only obvious
difference between the two profiles is the emission at 7300~\kms ~in the GBT profile that is not recovered in the VLBA
profile. This emission might be either produced by many \water ~maser features below our detection threshold of $3\sigma$ or
completely resolved out in the VLBI observations. \\
\begin{figure}[h]
\centering
\includegraphics[width = 9 cm]{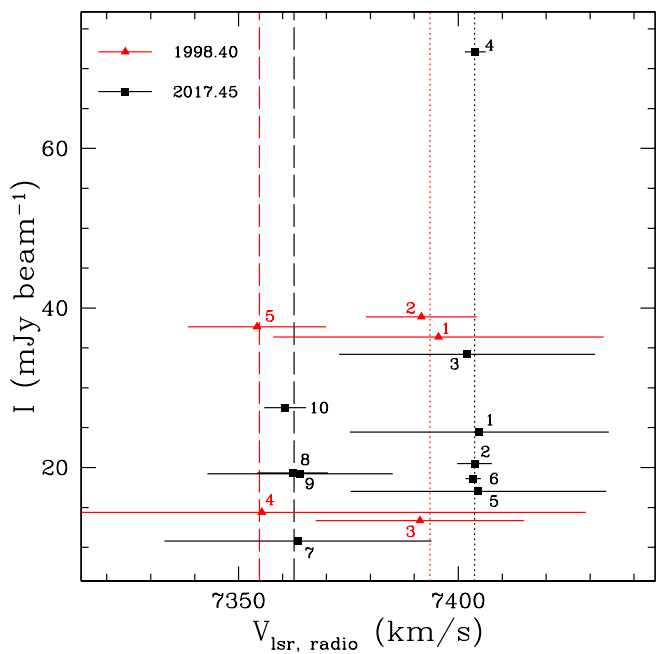}
\caption{Plot of the peak flux densities vs the peak velocities of the Gaussian components fitted to the \water 
~maser features detected in epochs 1998.40 (red triangles) and 2017.45 (black squares). The segments represent the 
$\Delta v\rm{_{L}}$ of the Gaussian components. The dashed and dotted lines indicate the mean velocities of the 
north (TXS.B03 and TXS.V04--V05) and the central (TXS.B01 and TXS.V01--V03) groups of the \water ~maser features,
respectively, at the two different epochs. The two components (3 and 4) of TXS.B02 have been considered be part of the
intermediate gas between the two groups and therefore they have been separated for the calculation of the mean velocities.
The numbers correspond to the fitted Gaussian components as reported in Tables~\ref{1998_tab} and \ref{2017_tab}. Refer to
Sect.~\ref{comp} for more details.} 
\label{velplot}
\end{figure}
\begin{figure*}[th!]
\centering
\includegraphics[width = 18.5 cm]{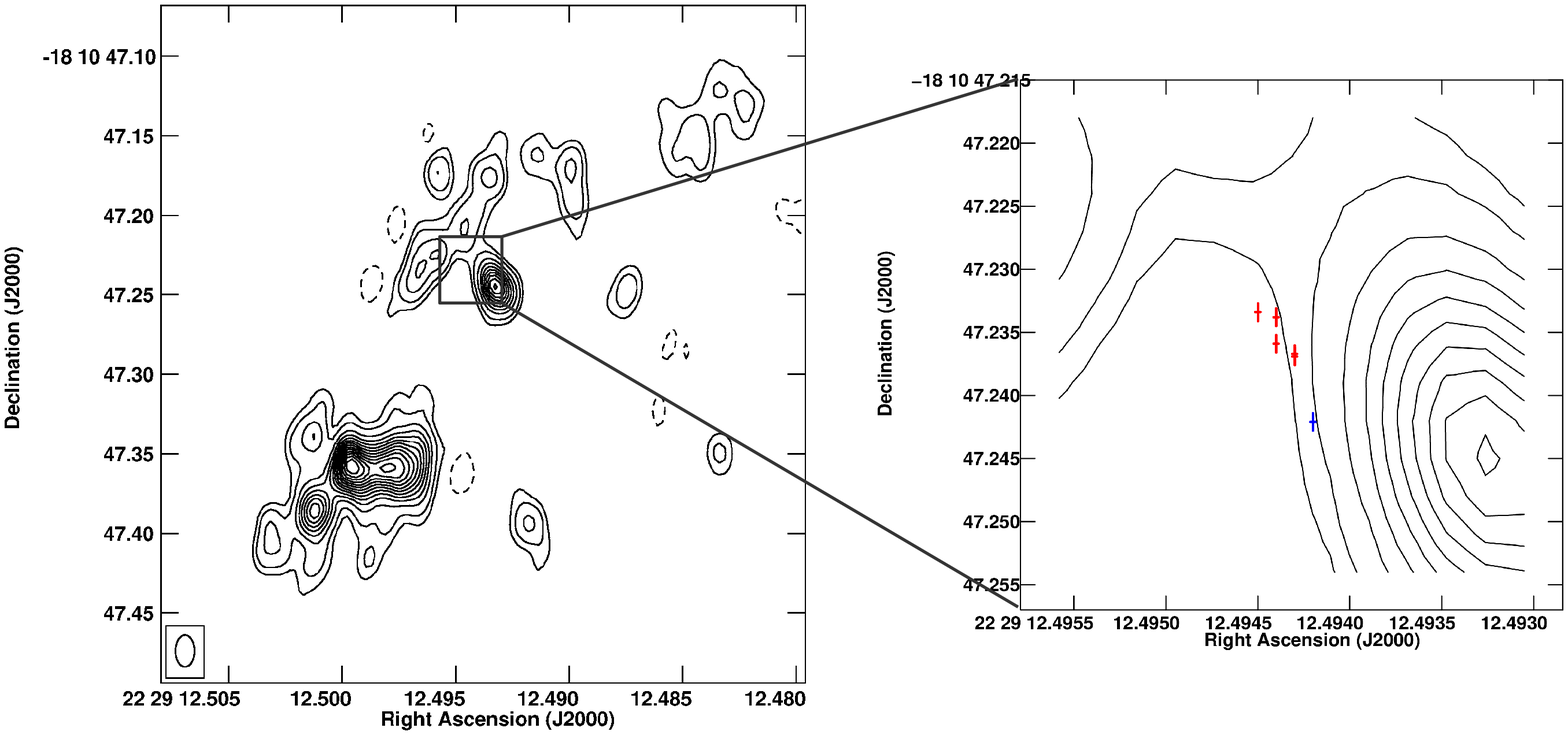}
\caption{Continuum emission of 1.4~GHz from the nuclear region of \txs ~as observed with the VLBA in 2002 \citep{tay04}.
Contours are $-1, 1, 2, 4, 6, 8, ..., 26 \times3\sigma$, where $\sigma=43.23~\rm{\mu Jy ~beam^{-1}}$. Red and blue
crosses indicate the positions of the red-shifted and blue-shifted \water ~masers with respect to the systemic velocity of
the galaxy ($V_{\rm{lsr,radio}}^{\rm{TXS}}=7270$~\kms), respectively, detected with the VLBA in epoch 2017.45 (see
Sect.~\ref{2017_res}). The size of the crosses correspond to the absolute position errors.}
\label{taylor}
\end{figure*}
\indent Thanks to the similar beam size and spectral resolution we can now compare the two VLBA epochs
1998.40 and 2017.45. The number of \water ~maser features detected in epoch 2017.45 is double than that in epoch 1998.40,
despite the total luminosity is comparable. Note that the difference of spectral line noise at the two epochs is only of
1.5~\mjybksl ~(see Tables~\ref{1998_tab} and \ref{2017_tab}). This suggests that the maser activity after about
20~years has not changed significantly, as seen from the single-dish observations (e.g., \citealt{bra03}). 
At both epochs the maser features are linearly distributed with the same position angle 
($\rm{PA_{1998.40}}=+26$\d$\pm16$\d ~and $\rm{PA_{2017.45}}=+28$\d$\pm12$\d). A blue-shifted maser feature is
detected only in epoch 2017.45 and its luminosity ($\sim80~L_{\odot}$) is still much larger than the nominal threshold
(10~$L_{\odot}$) used to classify an extragalactic \water ~maser as kilomaser or megamaser (see Sect~\ref{intro}). 
Although we did not 
\begin{figure}[h]
\centering
\includegraphics[width = 8 cm]{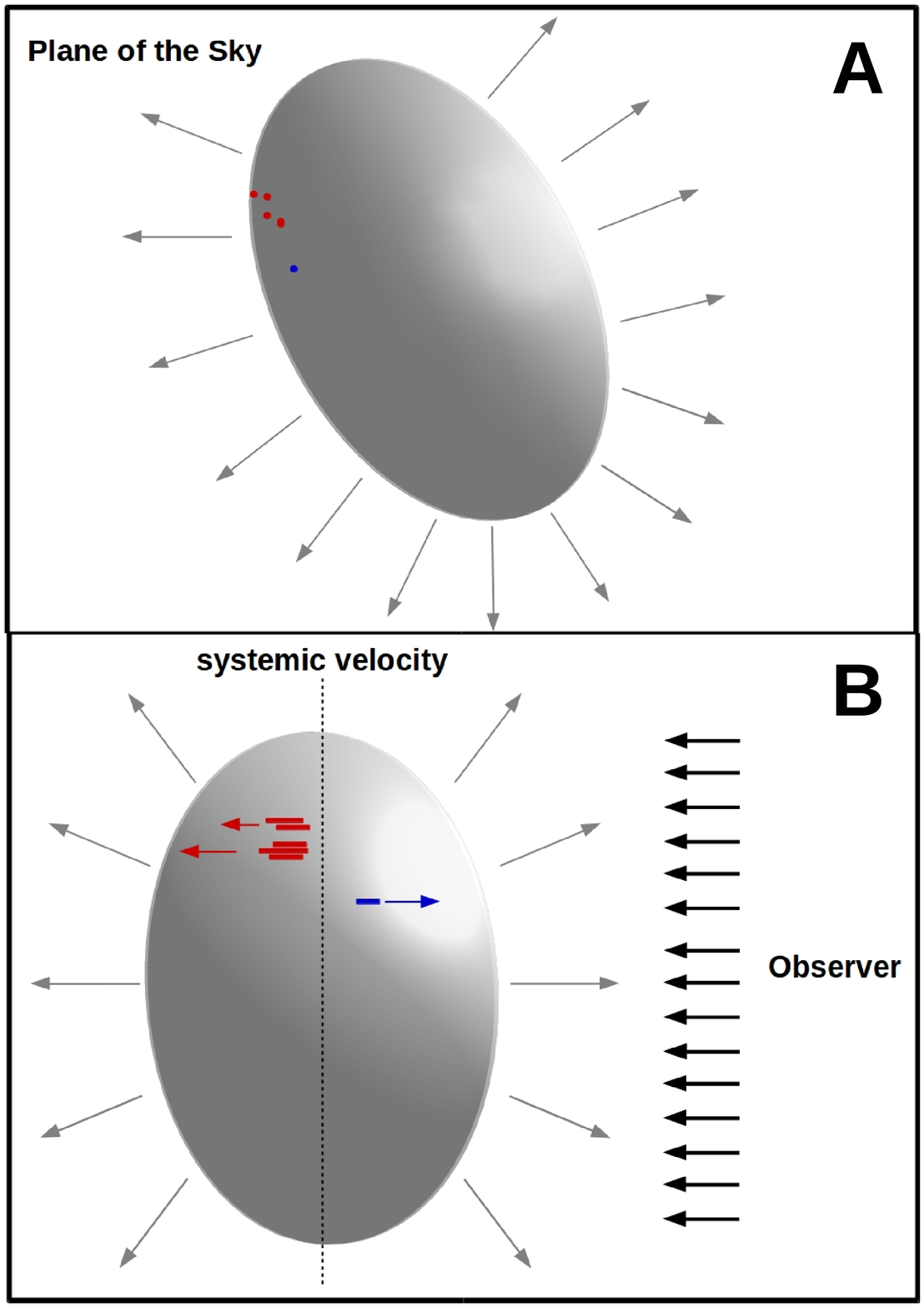}
\caption{Drawing sketch of the proposed location of the \water ~maser features. The gray ellipsoid represents the
molecular gas where the maser emission originates, the gray arrows around the ellipsoid represent the expansion directions of
the shock that inverted the population of the maser levels. The top panel (A) shows the location of the maser features as
projected on the plane of the sky (see Fig.~\ref{taylor}). The red dots are the red-shifted maser features (TXS.V01 --
TXS.V05) with respect to the systemic velocity of the galaxy, the blue dot is the blue-shifted maser feature (TXS.V06). The
bottom panel (B) shows the proposed location of the maser features as seen from east, therefore the point of view of the
observer is on the right of the image (black arrows). The red segments are the amplification paths of the red-shifted maser
feature (TXS.V01 -- TXS.V05), the blue segment is the amplification path of the blue-shifted maser feature (TXS.V06). The 
length of the segments are not physical, the different length indicates the different peak flux density (see
Table~\ref{vlba_evn}). The size and versus of the red and blue arrows (pointing towards left and right, respectively) indicate
the velocity difference between the maser features and the systemic velocity (dotted vertical line) of the galaxy.}
\label{3d}
\end{figure}
detect this blue-shifted maser feature in epoch 1998.40, its relative position with respect to the other red-shifted maser
features coincides with the position indicated by \citet{bal05}.\\
\indent The morphology of the line profiles between the two epochs are quite different, even though almost every maser
line is fitted by two Gaussian components. In epoch 1998.40 the Gaussian components are all broader than in epoch
2017.4. In particular $\Delta v\rm{_{L}}$ of the broadest components of the maser features are on average double
than those of the broadest components in epoch 2017.45, and $\Delta v\rm{_{L}}$ of the narrowest components of the maser
features are on average four times larger than those of the narrowest components in epoch 2017.45. Therefore the maser
features appear on average to become narrower. Considering the ratio between the
linewidth of the two components within the same maser feature, we have that in epoch 1998.40 there is a factor of
three between the narrow and the broad components, while in epoch 2017.45 this ratio is in the range between four and 17.
In particular, the narrow components in epoch 2017.45 show a particularly small linewidth of the order of a few \kms. The
difference in linewidth between the two epochs might be due either to a further amplification of the maser emission (e.g.,
increment of the number of \water ~molecules with a population inversion) but still remaining in the unsaturated regime, or
to the characteristic of the gas that the maser features are tracing (e.g., \citealt{gray}).\\
\indent Another aspect that deserves to be discussed is the difference of the peak velocities of the Gaussian components
between the two epochs. This difference is notable by comparing Table~\ref{1998_tab} and Table~\ref{2017_tab}, but it is
more evident from Fig.~\ref{velplot}. Here, we plot the peak flux densities of the Gaussian components as function of their 
peak velocities for the two epochs (red triangles for epoch 1998.40 and black squares for epoch 2017.45). The horizontal
bars indicate the linewidth ($\Delta v\rm{_{L}}$) of the Gaussian components and the vertical lines the mean velocity of
each maser group at the two epochs (red for epoch 1998.40 and black for epoch 2017.45). We grouped the maser features as
follow: TXS.B01 and TXS.V01--V03
are part of the central group and TXS.B03 and TXS.V04--V05 are part of the north group. The two Gaussian components of
feature TXS.B02 show a peak velocity difference of $\sim40$~\kms, therefore for the calculation of the mean velocities we
split them between the two groups according to their peak velocity (component 4 in the north group and component 3 in the
central group). The blue-shifted maser feature TXS.V06 is not considered here. The mean velocities are: for the north
group (dashed lines) 7354.8~\kms ~and 7362.6~\kms ~in epochs 1998.40 and 2017.45, respectively; and 7393.6~\kms ~and
7403.7~\kms ~for the central group (dotted lines) in epochs 1998.40 and 2017.45, respectively. The velocity difference 
is $+7.8$~\kms ~and $+10.1$~\kms ~for the north and central groups, respectively, suggesting a possible
acceleration of the gas in the last 20~years or, more likely, maser variability. That is, the brightest maser features
observed in 2017.45 were weaker in 1998.40 and viceversa. This might explain the relative prominence of the individual
features at the two different epochs. A systematic "acceleration" caused by a technical problem in data acquisition and/or
during the data reduction is unlikely because of the large number of spectral channels involved ($\sim40$).
\subsection{The nature of the (giga)maser}
\label{nature}
\citet{tay04} presented the 1.4~GHz radio continuum emission map of the nuclear region of \txs ~at a resolution of 
$20 \times 12$~mas (the phase-reference calibrator was J2236-1706 at a distance from \txs ~of $1^\circ\!.98$).
This shows several clumps of emission that they supposed to originate from a jet extending over 100~pc with a
$\rm{PA_{outflow}}=-36$\d.
They discussed the nature of the gigamaser in \txs ~by considering three different scenarios without knowing the exact
position of the maser emission, but favoring the jet nature of the \water ~maser. For the first time we are able to
provide the absolute position of the maser emission and, more important, to compare the position of the maser 
distribution with the continuum emission observed by \citet{tay04} with a relative accuracy of 10~mas (assuming position
errors equal of half-beam of the continuum emission since no position errors were reported in the paper).  The \water ~maser features detected with the VLBA in epoch 2017.45 are
overplotted to the 1.4~GHz radio continuum emission in
Fig.~\ref{taylor}, the error bars of the absolute positions are also shown. The \water
~maser emission originates close to the most luminous north-clump, in particular the red-shifted maser features
apparently follow very well the morphology of the radio continuum emission at 3$\sigma$ and the blue-shifted maser feature 
the emission at 6$\sigma$. \citet{tay04} measured towards that clump \hi ~absorption at the systemic velocity of
\txs ~with an optical depth in the interval $0.4<\tau<0.6$. From their Fig.~4 and knowing the masers position, we find that
the maser emission is associated to the gas with the largest $\tau$, as one may expect.\\
\indent From standard maser theory, the population inversion in the \water ~molecules, that is necessary for producing the
maser emission at 22~GHz, is
due to the passage of a shock through the molecular gas (e.g., \citealt{gray}). The 22~GHz \water ~masers are therefore
detected in the post-shocked gas (e.g., \citealt{god17}). Because in \txs ~the maser features are located at the edge of the
3$\sigma$ continuum emission, it is possible that the shock originates from the continuum emission and expands outwards (see
panel A of Fig.~\ref{3d}). Taking into account the peak flux density of the maser features (Col.~5 of Table~\ref{vlba_evn})
and the velocity difference along the line of sight between the velocities of the maser features (Col.~6 of 
Table~\ref{vlba_evn}) and the velocity of the nuclear region of \txs, which we assume to be equal to the recessional
velocity of the galaxy (7270~\kms), we can speculatively try to locate the masers in the 3D space (panel B of
Fig.~\ref{3d}). We represent the molecular gas where the maser emission arises as a regular ellipsoid, i.e., with the same
section from all lateral points of view. The ellipsoid is larger than the 3$\sigma$ contour of the continuum emission
because we assume that the gas extends at least up to where the maser features arise. We also assumed that the axis of the
ellipsoid moves at the velocity of the nuclear region of \txs, therefore the red-shifted maser features are located farther away
from the
observer than the blue-shifted one. Following this, we sketch the possible locations of the amplification paths of the maser
features in panel~B of Fig.~\ref{3d}; we do not have any indication on how deep they are into the ellipsoid though. We
assumed in Fig.~\ref{3d} that the brighter the maser feature is, the more embedded into the ellipsoid it is because of the
longer maser amplification path. Nevertheless,
the length of the segments in panel~B of Fig.~\ref{3d} do not represent the real length of the maser amplification paths,
because any estimation of those is impossible with the information we have at the moment, but it helps the reader to
visualize our idea. Of course this is one of the possible explanations of the pumping of the maser features. Indeed, it might
be plausible as well that the shock does not have any relation with the continuum emission. In this case the shock might
originate somewhere on the southeast of the most luminous north-clump and moves towards it, i.e., the shock shapes the
continuum emission but it is not part of it.\\
\indent Although we were able to finally determine the absolute position of the \water ~maser features in \txs, the position
of the nucleus of the galaxy is still uncertain. In the literature it is reported that the most luminous 22~GHz \water
~masers (the megamasers) are associated with AGNs, either to their accretion disk (called disk-type maser; e.g., NGC\,4258,
\citealt{miy95}) or, less common (possibly because less investigated), to the jet (jet-type maser; e.g., Mrk\,348,
\citealt{pec03}) or to the outflow (e.g., Circinus; \citealt{gre03}).
In both types the masers could show on the plane of the sky a
linear distribution, but what mainly differ them, apart the relative orientation with respect to the continuum emission
produced by the disk or the jet, is the single-dish line profile (e.g., \citealt{tar12}). In the disk-type profile three 
groups of narrow
lines ($<10$~\kms; e.g., NGC\,4258, \citealt{hum13}, and references therein) are observed: one at the systemic velocity,
and one each blue- and red-shifted from this. Generally, the velocity separation between the systemic lines to the others
is of the order of hundreds \kms ~(e.g., \citealt{pes15}). However, one or two groups of them might be undetected (e.g., 
NGC\,4388, \citealt{kuo11}). For the jet-type maser only one group of lines
is observed and usually they are much wider than the disk-type lines (e.g., \citealt{gal01,hen05}). Therefore we cannot
determine where the center of the AGN is located from the maser emission. What we can do instead is to discuss the two most
plausible scenarios.\\
\indent Following \citet{tay04}, if the continuum emission is due to a two-sided jet then we can assume that the
AGN (i.e., the black hole) is located in the gap between the two largest continuum emissions (case I). In this case the
maser features cannot trace
the disk, but they trace a shock due to the jet (jet-like maser). This shock can be either the one we discussed above and
showed in  Fig.~\ref{3d}, which we can think to be produced by a "bubble" that originates from the jet, or to the shock that
hits the gas from southeast of the north clump. Alternatively, if the black hole is located where the optical depth is
larger, therefore in the most luminous north clump \citep{tay04} (case II), two options are viable: the black hole
is at the center of the continuum emission (case IIa) or the black hole is where the \water ~maser features arise 
(case IIb). In case IIa the ellipsoid of Fig.~\ref{3d} is the jet/outflow that is ejected from the black hole and
consequently our idea described above agrees and we have a new case of jet-type maser. In case IIb the maser features might
trace the disk as suggested by \citet{bal05} and we might have a disk-type maser. 
In this later case, our Fig.~\ref{3d} does not apply. From the data at our disposal it is also difficult to
assess if the entire radio continuum emission can be associated to a radio jet and not, partly, to star formation. 
Indeed, LINER are often believed to host prominent nuclear star formation activity. However, the \water ~maser features in
\txs ~are so luminous that they can be confidently associated with the AGN activity.\\
\indent One could think that the profile of the \water ~maser features could help to disentangle the nature of the
maser. Actually, this adds more uncertainty. The single dish profile of the \water ~maser (Fig.~\ref{GBT}) 
shows only one group of maser features located around ($200$~\kms) the systemic velocity of \txs ~and does not clearly show
the other two satellite groups, of course this might be due to the weakness of the other two satellite groups. 
On the other side, the single-dish profile shows a notable stability over a years-range (e.g., \citealt{bra03}), 
that is hardly found in jet/outflow-associated masers. Indeed, these kind of masers usually display (highly) variable features
(\citealt{tar12}, and references therein). The one in \txs ~seems to resemble more closely the case of the \water ~maser found
in the gravitationally-lensed Quasar MGJ0414+0534 \citep{imp08}, whose nature is still disputed, for which a monthly-cadence 
monitoring of the broad maser line has not shown significant flux-density changes within a 15-months period \citep{cas11}. 
From the VLBI
profile we see for almost each maser feature two components: one very narrow, with $\Delta v\rm{_{L}}<10$~\kms ~ that is
typical of disk-type maser, and one very broad, with $\Delta v\rm{_{L}}>20$~\kms ~typical of jet-type maser (see
Table~\ref{vlba_evn}). Note that with the present data we cannot infer any significant consideration on the acceleration of
the maser features due to a Keplerian rotation. From the lines profile and the spatial distribution of the maser features it
is very difficult to determine the nature of the 22~GHz \water ~maser emission, a combination of the two types might also be
the case. However, further observations of different tracers of outflow/jet material at comparable spatial resolution of the
VLBI and a better knowledge of the radio continuum in the nucleus (e.g., spectral index) might help to answer the several
fundamental questions regarding \txs, in particular where the black hole of the AGN and its jet are.\\
\section{Summary}
The 22~GHz \water ~gigamaser in the galaxy \txs ~was observed twice in phase-reference mode and one in full polarization mode,
for a total of 3 epochs, with the VLBA and the EVN. We detected six \water ~maser features and thanks to the measurements of
the absolute positions we were able to associate them with the most luminous radio continuum clump of the nuclear region of the
galaxy. But what the maser features actually trace is still uncertain. Indeed, they might trace either a jet/outflow or the
accretion disk of the AGN located in the nuclear region of \txs. With the data at our disposal we also cannot completely
rule-out a possible association with star formation, even though the extreme high luminosity of the maser features seems to be
sufficient to exclude it. Furthermore, no polarized maser emission has been detected. Further observations, especially radio
continuum observations at different frequencies (e.g., to determine spectral indexes), are necessary to disentangle the possible
origins of the 22~GHz \water ~maser in \txs.
\begin{acknowledgements}
We wish to thank the anonymous referee for the useful suggestions that have improved the paper. 
The European VLBI Network is a joint facility of independent European, African, Asian, and North American radio 
astronomy institutes. Scientific results from data presented in this publication are derived from the following EVN 
project code(s): ES084. We thank G.B.~Taylor for providing the FITS file we used to make Fig.~\ref{taylor}.
\end{acknowledgements}

\bibliographystyle{aa}
\bibliography{biblio}

\begin{thebibliography}{43}
\expandafter\ifx\csname natexlab\endcsname\relax\def\natexlab#1{#1}\fi

\bibitem[{{Ball} {et~al.}(2005){Ball}, {Greenhill}, {Moran}, {Zaw}, \&
  {Henkel}}]{bal05}
{Ball}, G.~H., {Greenhill}, L.~J., {Moran}, J.~M., {Zaw}, I., \& {Henkel}, C.
  2005, in Astronomical Society of the Pacific Conference Series, Vol. 340,
  Future Directions in High Resolution Astronomy, ed. J.~{Romney} \& M.~{Reid},
  235

\bibitem[{{Barvainis} \& {Antonucci}(2005)}]{bar05}
{Barvainis}, R. \& {Antonucci}, R. 2005, \apjl, 628, L89

\bibitem[{{Bennert} {et~al.}(2009){Bennert}, {Barvainis}, {Henkel}, \&
  {Antonucci}}]{ben09}
{Bennert}, N., {Barvainis}, R., {Henkel}, C., \& {Antonucci}, R. 2009, \apj,
  695, 276

\bibitem[{{Bennert} {et~al.}(2004){Bennert}, {Schulz}, \& {Henkel}}]{ben04}
{Bennert}, N., {Schulz}, H., \& {Henkel}, C. 2004, \aap, 419, 127

\bibitem[{{Braatz} {et~al.}(2018){Braatz}, {Condon}, {Henkel}, {Greene}, {Lo},
  {Reid}, {Pesce}, {Gao}, {Impellizzeri}, {Kuo}, {Zhao}, {Constantin}, {Hao},
  \& {Litzinger}}]{bra18}
{Braatz}, J., {Condon}, J., {Henkel}, C., {et~al.} 2018, in IAU Symposium, Vol.
  336, Astrophysical Masers: Unlocking the Mysteries of the Universe, ed.
  A.~{Tarchi}, M.~J. {Reid}, \& P.~{Castangia}, 86--91

\bibitem[{{Braatz} {et~al.}(1994){Braatz}, {Wilson}, \& {Henkel}}]{bra94}
{Braatz}, J.~A., {Wilson}, A.~S., \& {Henkel}, C. 1994, \apjl, 437, L99

\bibitem[{{Braatz} {et~al.}(2003){Braatz}, {Wilson}, {Henkel}, {Gough}, \&
  {Sinclair}}]{bra03}
{Braatz}, J.~A., {Wilson}, A.~S., {Henkel}, C., {Gough}, R., \& {Sinclair}, M.
  2003, \apjs, 146, 249

\bibitem[{{Burns} {et~al.}(2019){Burns}, {Orosz}, {Bayandina}, {Surcis},
  {Olech}, {MacLeod}, {Volvach}, {Rudnitskii}, {Hirota}, {Blanchard},
  {Marcote}, {van Langevelde}, {Chibueze}, {Sugiyama}, {Kim}, {Val'tts},
  {Shakhvorostova}, {Kramer}, {Baan}, {Brogan}, {Hunter}, {Kurtz}, {Sobolev},
  {Brand}, \& {Volvach}}]{bur19}
{Burns}, R.~A., {Orosz}, G., {Bayandina}, O., {et~al.} 2019, \mnras, 2756

\bibitem[{{Castangia} {et~al.}(2011){Castangia}, {Impellizzeri}, {McKean},
  {Henkel}, {Brunthaler}, {Roy}, {Wucknitz}, {Ott}, \& {Momjian}}]{cas11}
{Castangia}, P., {Impellizzeri}, C.~M.~V., {McKean}, J.~P., {et~al.} 2011,
  \aap, 529, A150

\bibitem[{{Castangia} {et~al.}(2019){Castangia}, {Surcis}, {Tarchi},
  {Caccianiga}, {Severgnini}, \& {Della Ceca}}]{cas19}
{Castangia}, P., {Surcis}, G., {Tarchi}, A., {et~al.} 2019, \aap, 629, A25

\bibitem[{{Castangia} {et~al.}(2016){Castangia}, {Tarchi}, {Caccianiga},
  {Severgnini}, \& {Della Ceca}}]{cas16}
{Castangia}, P., {Tarchi}, A., {Caccianiga}, A., {Severgnini}, P., \& {Della
  Ceca}, R. 2016, \aap, 586, A89

\bibitem[{{Claussen} {et~al.}(1998){Claussen}, {Diamond}, {Braatz}, {Wilson},
  \& {Henkel}}]{cla98}
{Claussen}, M.~J., {Diamond}, P.~J., {Braatz}, J.~A., {Wilson}, A.~S., \&
  {Henkel}, C. 1998, \apjl, 500, L129

\bibitem[{{Falcke} {et~al.}(2000){Falcke}, {Wilson}, {Henkel}, {Brunthaler}, \&
  {Braatz}}]{fal00}
{Falcke}, H., {Wilson}, A.~S., {Henkel}, C., {Brunthaler}, A., \& {Braatz},
  J.~A. 2000, \apjl, 530, L13

\bibitem[{{Gallimore} {et~al.}(2001){Gallimore}, {Henkel}, {Baum}, {Glass},
  {Claussen}, {Prieto}, \& {Von Kap-herr}}]{gal01}
{Gallimore}, J.~F., {Henkel}, C., {Baum}, S.~A., {et~al.} 2001, \apj, 556, 694

\bibitem[{{Gao} {et~al.}(2017){Gao}, {Braatz}, {Reid}, {Condon}, {Greene},
  {Henkel}, {Impellizzeri}, {Lo}, {Kuo}, {Pesce}, {Wagner}, \& {Zhao}}]{gao17}
{Gao}, F., {Braatz}, J.~A., {Reid}, M.~J., {et~al.} 2017, \apj, 834, 52

\bibitem[{{Goddi} {et~al.}(2017){Goddi}, {Surcis}, {Moscadelli}, {Imai},
  {Vlemmings}, {van Langevelde}, \& {Sanna}}]{god17}
{Goddi}, C., {Surcis}, G., {Moscadelli}, L., {et~al.} 2017, \aap, 597, A43

\bibitem[{{Gray}(2012)}]{gray}
{Gray}, M. 2012, {Maser Sources in Astrophysics}

\bibitem[{{Greenhill} {et~al.}(2003){Greenhill}, {Booth}, {Ellingsen},
  {Herrnstein}, {Jauncey}, {McCulloch}, {Moran}, {Norris}, {Reynolds}, \&
  {Tzioumis}}]{gre03}
{Greenhill}, L.~J., {Booth}, R.~S., {Ellingsen}, S.~P., {et~al.} 2003, \apj,
  590, 162

\bibitem[{{Henkel} {et~al.}(2018){Henkel}, {Greene}, \& {Kamali}}]{hen18}
{Henkel}, C., {Greene}, J.~E., \& {Kamali}, F. 2018, in IAU Symposium, Vol.
  336, Astrophysical Masers: Unlocking the Mysteries of the Universe, ed.
  A.~{Tarchi}, M.~J. {Reid}, \& P.~{Castangia}, 69--79

\bibitem[{{Henkel} {et~al.}(2005){Henkel}, {Peck}, {Tarchi}, {Nagar}, {Braatz},
  {Castangia}, \& {Moscadelli}}]{hen05}
{Henkel}, C., {Peck}, A.~B., {Tarchi}, A., {et~al.} 2005, \aap, 436, 75

\bibitem[{{Humphreys} {et~al.}(2013){Humphreys}, {Reid}, {Moran}, {Greenhill},
  \& {Argon}}]{hum13}
{Humphreys}, E.~M.~L., {Reid}, M.~J., {Moran}, J.~M., {Greenhill}, L.~J., \&
  {Argon}, A.~L. 2013, \apj, 775, 13

\bibitem[{{Impellizzeri} {et~al.}(2008){Impellizzeri}, {McKean}, {Castangia},
  {Roy}, {Henkel}, {Brunthaler}, \& {Wucknitz}}]{imp08}
{Impellizzeri}, C.~M.~V., {McKean}, J.~P., {Castangia}, P., {et~al.} 2008,
  \nat, 456, 927

\bibitem[{{Keimpema} {et~al.}(2015){Keimpema}, {Kettenis}, {Pogrebenko},
  {Campbell}, {Cim{\'o}}, {Duev}, {Eldering}, {Kruithof}, {van Langevelde},
  {Marchal}, {Molera Calv{\'e}s}, {Ozdemir}, {Paragi}, {Pidopryhora},
  {Szomoru}, \& {Yang}}]{kei15}
{Keimpema}, A., {Kettenis}, M.~M., {Pogrebenko}, S.~V., {et~al.} 2015,
  Experimental Astronomy, 39, 259

\bibitem[{{Koekemoer} {et~al.}(1995){Koekemoer}, {Henkel}, {Greenhill}, {Dey},
  {van Breugel}, {Codella}, \& {Antonucci}}]{koe95}
{Koekemoer}, A.~M., {Henkel}, C., {Greenhill}, L.~J., {et~al.} 1995, \nat, 378,
  697

\bibitem[{{Kuo} {et~al.}(2011){Kuo}, {Braatz}, {Condon}, {Impellizzeri}, {Lo},
  {Zaw}, {Schenker}, {Henkel}, {Reid}, \& {Greene}}]{kuo11}
{Kuo}, C.~Y., {Braatz}, J.~A., {Condon}, J.~J., {et~al.} 2011, \apj, 727, 20

\bibitem[{{Kuo} {et~al.}(2018){Kuo}, {Constantin}, {Braatz}, {Chung},
  {Witherspoon}, {Pesce}, {Impellizzeri}, {Gao}, {Hao}, {Woo}, \&
  {Zaw}}]{kuo18}
{Kuo}, C.~Y., {Constantin}, A., {Braatz}, J.~A., {et~al.} 2018, \apj, 860, 169

\bibitem[{{McCallum} {et~al.}(2007){McCallum}, {Ellingsen}, \&
  {Lovell}}]{mcc07}
{McCallum}, J.~N., {Ellingsen}, S.~P., \& {Lovell}, J.~E.~J. 2007, \mnras, 376,
  549

\bibitem[{{Miyoshi} {et~al.}(1995){Miyoshi}, {Moran}, {Herrnstein},
  {Greenhill}, {Nakai}, {Diamond}, \& {Inoue}}]{miy95}
{Miyoshi}, M., {Moran}, J., {Herrnstein}, J., {et~al.} 1995, \nat, 373, 127

\bibitem[{{Modjaz} {et~al.}(2005){Modjaz}, {Moran}, {Kondratko}, \&
  {Greenhill}}]{mod05}
{Modjaz}, M., {Moran}, J.~M., {Kondratko}, P.~T., \& {Greenhill}, L.~J. 2005,
  \apj, 626, 104

\bibitem[{{Nedoluha} \& {Watson}(1992)}]{ned92}
{Nedoluha}, G.~E. \& {Watson}, W.~D. 1992, \apj, 384, 185

\bibitem[{{Peck} {et~al.}(2003){Peck}, {Henkel}, {Ulvestad}, {Brunthaler},
  {Falcke}, {Elitzur}, {Menten}, \& {Gallimore}}]{pec03}
{Peck}, A.~B., {Henkel}, C., {Ulvestad}, J.~S., {et~al.} 2003, \apj, 590, 149

\bibitem[{{Pesce} {et~al.}(2015){Pesce}, {Braatz}, {Condon}, {Gao}, {Henkel},
  {Litzinger}, {Lo}, \& {Reid}}]{pes15}
{Pesce}, D.~W., {Braatz}, J.~A., {Condon}, J.~J., {et~al.} 2015, \apj, 810, 65

\bibitem[{{Reid} \& {Honma}(2014)}]{rei14}
{Reid}, M.~J. \& {Honma}, M. 2014, \araa, 52, 339

\bibitem[{{Sawada-Satoh} {et~al.}(2008){Sawada-Satoh}, {Kameno}, {Nakamura},
  {Namikawa}, {Shibata}, \& {Inoue}}]{saw08}
{Sawada-Satoh}, S., {Kameno}, S., {Nakamura}, K., {et~al.} 2008, \apj, 680, 191

\bibitem[{{Surcis} {et~al.}(2011){Surcis}, {Vlemmings}, {Curiel}, {Hutawarakorn
  Kramer}, {Torrelles}, \& {Sarma}}]{sur11}
{Surcis}, G., {Vlemmings}, W.~H.~T., {Curiel}, S., {et~al.} 2011, \aap, 527,
  A48

\bibitem[{{Tarchi}(2012)}]{tar12}
{Tarchi}, A. 2012, in IAU Symposium, Vol. 287, Cosmic Masers - from OH to H0,
  ed. R.~S. {Booth}, W.~H.~T. {Vlemmings}, \& E.~M.~L. {Humphreys}, 323--332

\bibitem[{{Tarchi} {et~al.}(2007){Tarchi}, {Brunthaler}, {Henkel}, {Menten},
  {Braatz}, \& {Wei{\ss}}}]{tar07}
{Tarchi}, A., {Brunthaler}, A., {Henkel}, C., {et~al.} 2007, \aap, 475, 497

\bibitem[{{Tarchi} {et~al.}(2011){Tarchi}, {Castangia}, {Henkel}, {Surcis}, \&
  {Menten}}]{tar11}
{Tarchi}, A., {Castangia}, P., {Henkel}, C., {Surcis}, G., \& {Menten}, K.~M.
  2011, \aap, 525, A91

\bibitem[{{Tarchi} {et~al.}(2003){Tarchi}, {Henkel}, {Chiaberge}, \&
  {Menten}}]{tar03}
{Tarchi}, A., {Henkel}, C., {Chiaberge}, M., \& {Menten}, K.~M. 2003, \aap,
  407, L33

\bibitem[{{Taylor} {et~al.}(2002){Taylor}, {Peck}, {Henkel}, {Falcke},
  {Mundell}, {O'Dea}, {Baum}, \& {Gallimore}}]{tay02}
{Taylor}, G.~B., {Peck}, A.~B., {Henkel}, C., {et~al.} 2002, \apj, 574, 88

\bibitem[{{Taylor} {et~al.}(2004){Taylor}, {Peck}, {Ulvestad}, \&
  {O'Dea}}]{tay04}
{Taylor}, G.~B., {Peck}, A.~B., {Ulvestad}, J.~S., \& {O'Dea}, C.~P. 2004,
  \apj, 612, 780

\bibitem[{{Vlemmings} {et~al.}(2007){Vlemmings}, {Bignall}, \&
  {Diamond}}]{vle07}
{Vlemmings}, W.~H.~T., {Bignall}, H.~E., \& {Diamond}, P.~J. 2007, \apj, 656,
  198

\bibitem[{{Zhao} {et~al.}(2018){Zhao}, {Braatz}, {Condon}, {Lo}, {Reid},
  {Henkel}, {Pesce}, {Greene}, {Gao}, {Kuo}, \& {Impellizzeri}}]{zha18}
{Zhao}, W., {Braatz}, J.~A., {Condon}, J.~J., {et~al.} 2018, \apj, 854, 124

\end{thebibliography}

\end{document}